\documentclass[journal,letter,10pt,twocolumn]{IEEEtran}
\pdfoutput=1

\usepackage{amsmath, amssymb, amscd, amsthm, amsfonts}
\usepackage{bbm}
\usepackage[pdftex]{graphicx}
\usepackage[caption=false,font=footnotesize]{subfig}
\usepackage{dblfloatfix}    
\usepackage{color}
\usepackage[linesnumbered,ruled,vlined]{algorithm2e}
\usepackage[noend]{algpseudocode}
\usepackage{comment}

\usepackage{hyperref}

\usepackage{tikz,ifthen}
\usetikzlibrary{positioning,arrows,backgrounds}
\usetikzlibrary{fit}
\usetikzlibrary{calc}
 \usepackage{transparent}

\usepackage{colortbl}

\newtheorem{lemma}{Lemma}
\newtheorem{corollary}{Corollary}
\newtheorem{proposition}{Proposition}
\newtheorem{definition}{Definition}

\usepackage{pgfplots}
\pgfplotsset{width=10cm,compat=1.9}

\usepackage{subfiles}
\usepackage[noadjust]{cite}

\usepackage[english]{babel}

\addto\captionsenglish{}

\graphicspath{{../Figures/}}



\usepackage[font=footnotesize]{caption}

\SetCommentSty{mycommfont}

\SetKwInput{KwInput}{Input}                
\SetKwInput{KwOutput}{Output}              

\begin{document}

\title{Capacity Bounds for One-Bit MIMO Gaussian \\ Channels with Analog Combining}
\author{
    \IEEEauthorblockN{Neil Irwin Bernardo, \textit{Graduate Student Member, IEEE}, Jingge Zhu, \textit{Member, IEEE}, Yonina C. Eldar, \textit{Fellow, IEEE}, and Jamie Evans, \textit{Senior Member, IEEE}
   }

    \thanks{Manuscript received April 8, 2022; revised July 28, 2022; accepted Sept 12, 2022. This project has received funding from the Australian Research Council under project DE210101497, from the European Research Council (ERC) under the European Union’s Horizon 2020 research and innovation programme (grant agreement No. 101000967), and from the Israel Science Foundation (grant No. 536/22). N.I. Bernardo acknowledges the Melbourne Research Scholarship of the University of Melbourne and the DOST-ERDT Faculty Development Fund of the Republic of the Philippines for sponsoring his doctoral studies. The associate editor coordinating the review of this article and approving it for publication was N. Liu. (\emph{Corresponding author: Neil Irwin Bernardo}.) }

    \thanks{N.I. Bernardo is with the Department of Electrical and Electronic Engineering, The University of Melbourne, Parkville, VIC 3010, Australia and also with the Electrical and Electronics Engineering Institute, University of the Philippines Diliman, Quezon City 1101, Philippines (e-mail: bernardon@student.unimelb.edu.au).}
    \thanks{J. Zhu and J. Evans are with the Department of Electrical and Electronic Engineering, The University of Melbourne, Parkville, VIC 3010, Australia (e-mail: jingge.zhu@unimelb.edu.au;
jse@unimelb.edu.au).}
      \thanks{Y. C. Eldar is with the Faculty of Math and CS, Weizmann Institute of Science, Rehovot 7610001, Israel (e-mail: yonina.eldar@weizmann.ac.il).}
}
\maketitle
\begin{abstract}
The use of 1-bit analog-to-digital converters (ADCs) is seen as a promising approach to significantly reduce the power consumption and hardware cost of multiple-input multiple-output (MIMO) receivers. However, the nonlinear distortion due to 1-bit quantization fundamentally changes the optimal communication strategy and also imposes a capacity penalty to the system. In this paper, the capacity of a Gaussian MIMO channel in which the antenna outputs are processed by an analog linear combiner and then quantized by a set of zero threshold ADCs is studied. A new capacity upper bound for the zero threshold case is established that is tighter than the bounds available in the literature. In addition, we propose an achievability scheme which configures the analog combiner to create parallel Gaussian channels with phase quantization at the output. Under this class of analog combiners, an algorithm is presented that identifies the analog combiner and input distribution that maximize the achievable rate. Numerical results are provided showing that the rate of the achievability scheme is tight in the low signal-to-noise ratio (SNR) regime. Finally, a new 1-bit MIMO receiver architecture which employs analog temporal and spatial processing is proposed. The proposed receiver attains the capacity in the high SNR regime.
\end{abstract}

\begin{IEEEkeywords}
MIMO Communications, Quantization, Analog Combining, Capacity 
\end{IEEEkeywords}

\IEEEpeerreviewmaketitle

\section{Introduction}\label{section-intro}

The use of multiple-input multiple-output (MIMO) technology has attracted considerable attention during the last two decades due to the significant capacity enhancement it offers. This is evident in the number of wireless broadband standards that have incorporated MIMO technology into their specifications. However, even though its theoretical gains are well-established in the literature, some practical concerns (e.g. hardware cost, power consumption) arise as more antenna elements are connected to the communication device \cite{Risi:2014}. Furthermore, recent advancements in next generation mobile technology have pushed forward the use of much wider transmission bandwidth and much larger antenna arrays to achieve its performance targets \cite{Rappaport:2013, Rangan:2014,Halbauer:2021}. Consequently, there is an increasing demand for new MIMO receiver designs that are energy-efficient and are able to \emph{reliably} support high data-rate applications.

High-speed and high-resolution analog-to-digital converters (ADCs) are one of the primary contributors to the power consumption of wireless receivers. The ADC power consumption scales linearly with the sampling rate and exponentially with the number of quantization bits per sample, regardless of the topology \cite{Walden:1999}. Thus, one straightforward design methodology in implementing low-power MIMO receivers is to simply replace the high-resolution ADCs connected to each radio frequency (RF) chain with low-resolution counterparts (usually 1-bit). We shall refer to this as the \emph{conventional} low-resolution ADC design. This receiver design is particularly attractive for massive MIMO systems due to the hardware scaling law -- that is, the impact of hardware imperfections on the overall spectral efficiency diminishes as the number of antenna elements grows \cite{Bjornson:2015}. In fact, some early results suggest that the power savings and cost reduction obtained from using conventional 1-bit ADC MIMO systems may outweigh the rate loss caused by severe quantization; even with simple linear detection schemes \cite{ Risi:2014,Jacobsson:2015,Li:2017}. Error rate analyses of single-input multiple-output (SIMO) fading channels with low-resolution output quantization have established that, under certain conditions, the diversity order of optimal detectors is nonzero and improves linearly with the number of receive antennas \cite{bernardo2021sep,Gayan:2021}. Furthermore, other receiver functionalities (e.g. timing recovery, channel estimation) have been shown to work with acceptable performance even if the receiver is equipped with low-resolution ADCs \cite{Sun:2010,Dabeer:2010,Liu:2018,Schluter:2020}.

The notable benefits of the conventional low-resolution receiver design have sparked interest in investigating the information-theoretic limits of communication channels with output quantization. The work of Singh et al. \cite{Singh:2009} appears to be the first to present exact results on the capacity of real additive white Gaussian noise (AWGN) channel with low-resolution ADCs. Several works that followed characterized the capacity-achieving input of various single-input single-output (SISO) channels with output quantization \cite{Mezghani:2007,Krone:2010, Koch:2013, Vu:2019, Rahman2:2020,bernardo2021TIT,bernardo2022polar}. Yet, while the main motivation of using low-resolution ADCs is for scalable implementation of massive MIMO systems, the aforementioned information-theoretic results have not been extended to the MIMO setting. To date, the capacity of quantized MIMO remains elusive and is analyzed through capacity bounds \cite{Mezghani:2012,Mo:2014, Orhan:2015,Demir:2021}. These bounds, however, can be loose in certain signal-to-noise ratio (SNR) regimes and problem settings. For instance, the tightness of the finite SNR upper bound and channel inversion lower bound established in \cite{Mo:2014} depends on the row rank and condition number of the channel. The additive quantization noise model (AQNM) capacity lower bound in \cite{Mezghani:2012,Orhan:2015} is also shown to be loose in the high SNR regime; thus hinting at the suboptimality of Gaussian signaling for quantized MIMO channels.

Recent literature surveys \cite{Zhang:2018,Liu:2019, Choi:2020} have examined new receiver architectures that are energy-efficient and cost-effective but incur less performance degradation than conventional low-resolution ADC systems. These architectures include the mixed-ADC receivers \cite{Liang:2016}, hybrid analog/digital receivers \cite{Mo:2017,Roth:2017,Shlezinger:2019,Shlezinger:2020}, and machine learning (ML)-based receivers \cite{Jeon:2018,Zhang:2020}. In \cite{Khalili:2021}, the authors suggest two new receiver designs, namely the \emph{hybrid blockwise receiver} and \emph{adaptive threshold receiver}, that perform analog spatial and temporal processing prior to 1-bit quantization. The idea of using analog linear combiners prior to 1-bit quantization, called the \emph{hybrid one-shot receiver}, was initially proposed in \cite{Rini:2017} as a means to compare performance of different MIMO systems under various output quantization constraints. Analog temporal processing was then incorporated in this quantized MIMO framework to demonstrate a fundamental tradeoff between latency and maximum achievable rate of quantized MIMO channels in the high SNR regime \cite{Khalili:2019}. 

In this paper, we first look at the performance of the hybrid one-shot receiver with zero threshold ADCs. In other words, we only observe the signs of the analog linear combiner outputs. The zero threshold ADC setup is expected to yield a lower capacity than its non-zero threshold counterpart. Yet, the zero threshold case is still interesting since it eliminates the need for an automatic gain control (AGC), which is otherwise required in multi-level quantization to match the dynamic range of the received signal \cite{Jacobsson:2017}. In addition, the hybrid one-shot receiver with zero threshold ADCs is essentially the 1-bit ADC version of the hybrid analog-digital acquisition system used for task-based quantization \cite{Shlezinger:2019,Shlezinger:2020}. We then introduce analog temporal processing to the receiver. A new MIMO receiver architecture is proposed which uses analog domain pipelining and adaptive phase shifters to attain higher achievable rate than the hybrid one-shot receiver with zero threshold ADCs. Our main contributions are summarized as follows:
\begin{itemize}
    \item We provide an achievability scheme for the hybrid one-shot receiver with zero threshold ADCs. The achievability scheme formulates the capacity problem as a nonconvex resource allocation problem. To this end, an alternating optimization approach is presented to obtain a local optimal solution to this nonconvex problem.
    \item Using the data processing inequality (DPI), we establish a new capacity upper bound in the finite SNR regime for the hybrid one-shot receiver with zero threshold ADCs. More precisely, by showing that the amplitude information is discarded in the zero threshold ADC case, we obtain an upper bound that is tighter than the truncated Shannon capacity (TSC) used in \cite{Khalili:2021}.
    \item Through numerical evaluation, we show that the output produced by our alternating optimization approach is tight in the low SNR regime. However, a gap between the capacity upper bound and achievability scheme exists in the high SNR regime. We characterize this gap as a function of the number of eigenchannels and number of sign quantizers.
    \item To close the gap mentioned above, we introduce a new ADC mechanism, called \emph{pipelined phase ADC}, which incorporates analog temporal processing in the quantization process. By incorporating this ADC mechanism to the receiver, we show that the high SNR capacity can be attained. We compare the achievable rate of our proposed receiver to that of the adaptive threshold receiver in \cite{Khalili:2021}.  Numerical results are presented showing that the proposed receiver outperforms the adaptive threshold receiver when the eigenvalues of the channel are equal.
    
\end{itemize}

The rest of the paper is organized as follows: Section \ref{section-problem_formulation} formulates the system model and the problem we aim to address. Sections \ref{section-achievability_scheme} and \ref{section-cap_upperbound} present the details of the achievability scheme and the derivation of the capacity upper bound, respectively, for the hybrid one-shot receiver with zero threshold ADCs. Section \ref{section-numerical_result1} provides numerical results and analysis for the achievability scheme and the capacity upper bound established in Sections \ref{section-achievability_scheme} and \ref{section-cap_upperbound}. Sections \ref{section-MIMO_phaseADC} and \ref{section-numerical_result2} discuss the proposed receiver along with analysis of its achievable rate. We also compare its performance with existing work. Finally, Section \ref{section-conclusion} concludes the paper.

\section{Problem Formulation}
\label{section-problem_formulation}

\begin{figure}[t]
    \centering
    \includegraphics[scale = .65]{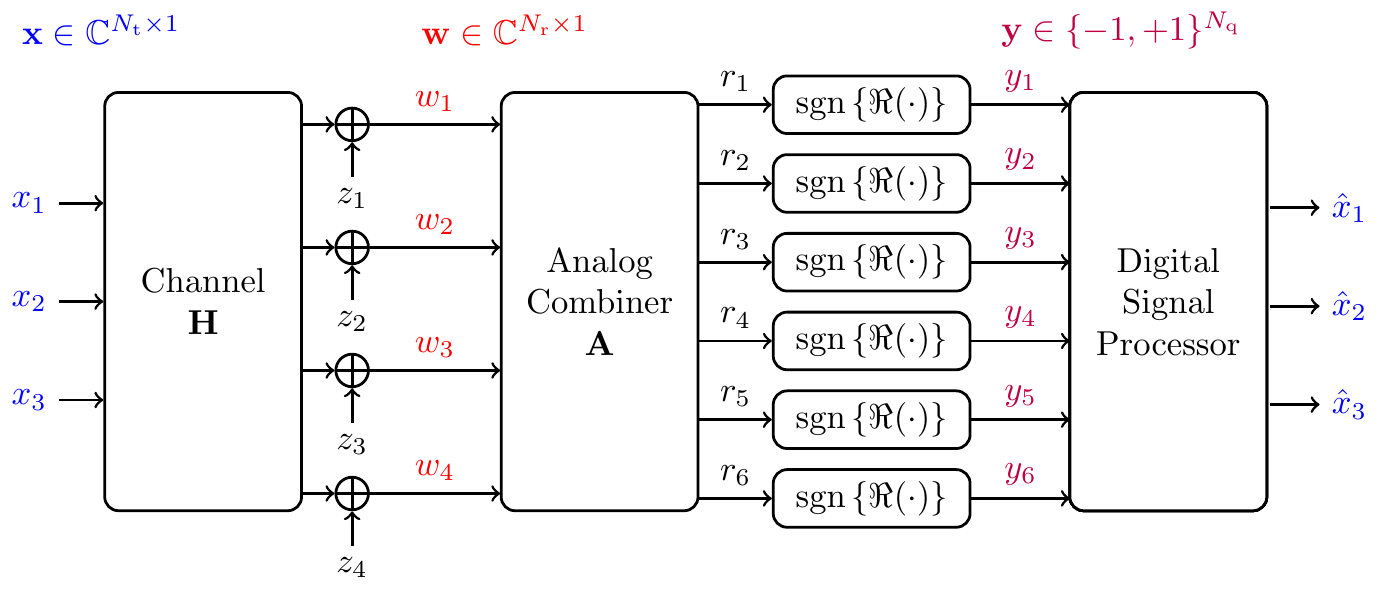}
    \caption{System Model of MIMO with Analog Combining prior to Output Quantization. Here, $N_\mathrm{t} = 3$, $N_\mathrm{r} = 4$, and $N_\mathrm{q} = 6$.}
    \label{fig:mimo_sys_model}
\end{figure}
We consider a discrete-time memoryless MIMO system shown in Figure \ref{fig:mimo_sys_model} with $N_\mathrm{t}$ transmit antennas and $N_\mathrm{r}$ receive antennas. The input vector $\mathbf{x}$ satisfies the power constraint $\mathbb{E}[||\mathbf{x}||^2] \leq P$. The input-output relationship between transmitted symbol $\mathbf{x} \in \mathbb{C}^{N_\mathrm{t}\times 1}$ and received symbol $\mathbf{w}  \in \mathbb{C}^{N_\mathrm{r}\times 1}$ is 
\begin{equation}\label{eq:sys_model_unquantized}
    \mathbf{w} =  \mathbf{H} \mathbf{x} +  \mathbf{z},
\end{equation}
where $\mathbf{H}\in\mathbb{C}^{N_\mathrm{r}\times N_\mathrm{t}}$ is the fixed MIMO channel gain known at the transmitter and receiver, and $\mathbf{z} \sim \mathcal{CN}(\boldsymbol{0},\sigma^2\boldsymbol{\mathrm{I}}_{N_\mathrm{r}})$ is a circular-symmetric zero-mean complex Gaussian noise vector with noise variance $\sigma^2$. We further assume that $\mathbf{H}$ is a full rank matrix. The receiver is equipped with $N_\mathrm{q}$ sign quantizers and an analog linear combiner $\mathbf{A} \in \mathbb{C}^{N_{\mathrm{q}}\times N_{\mathrm{r}}}$ preprocesses the signal before quantization. This receiver structure has been widely adopted in various applications, such as in task-based quantization \cite{Shlezinger:2019,Shlezinger:2020}. The output vector $\mathbf{y}\in \mathbb{C}^{N_\mathrm{q}\times 1}$ can be written as
\begin{align}\label{eq:sys_model_quantized}
    \mathbf{y} = \mathrm{sign}\left(\Re\{\mathbf{A}\mathbf{w}\}\right) = \mathrm{sign}\left(\Re\{\mathbf{A}\mathbf{H}\mathbf{x} + \mathbf{z}'\}\right),
\end{align}
where $\mathbf{z}' = \mathbf{A}\mathbf{z}$ and the $\mathrm{sign}\{\cdot\}$ function is applied to every element of the real vector.

We note the differences between our problem setup and the system model of the hybrid one-shot receiver formulated in \cite{Khalili:2018,Khalili:2021}. In their work, the quantized output vector $\mathbf{y}$ is expressed as $\mathbf{y} = \mathrm{sign}(\mathbf{A}\mathbf{w} + \mathbf{t})$, where $\mathbf{A}\in\mathbb{R}^{N_{\mathrm{q}}\times N_{\mathrm{r}}}$,$\mathbf{w}\in\mathbb{R}^{N_{\mathrm{r}}\times 1}$, and $\mathbf{t}\in\mathbb{R}^{N_{\mathrm{q}}\times 1}$. In contrast, the entries of $\mathbf{x}$, $\mathbf{H}$, and $\mathbf{A}$ in our problem setup are complex-valued quantities. The threshold vector $\mathbf{t}$ is also set to be the all-zero vector. The $\mathbf{t} = \mathbf{0}$ case is particularly appealing from a practical viewpoint since receivers employing zero-threshold ADCs do not require AGC during the data detection phase \cite{Risi:2014}; thus further reducing the hardware complexity and cost. Moreover, by restricting the setup to $\mathbf{t} = \mathbf{0}$, we are able to derive tighter capacity bounds for this problem setup than simply plugging in $\mathbf{t} = \mathbf{0}$ to the capacity bounds established in \cite{Khalili:2018,Khalili:2021} for general $\mathbf{t}$.

For a fixed configuration of the analog combiner $\mathbf{A}$, the capacity expression of model \eqref{eq:sys_model_quantized} can be written as
\begin{align}\label{eq:capacity_fixed_A}
    C(\mathbf{A}) = \underset{F_\mathbf{X}}{\max} \;I_{\mathbf{A}}(\mathbf{x};\mathbf{y}),
\end{align}
where we used the subscript $\mathbf{A}$ in (\ref{eq:capacity_fixed_A}) to indicate that the mutual information between the transmitted signals and the sign quantizer outputs are induced by a choice of $\mathbf{A}$. With slight abuse of notation, we use $F_{\mathbf{X}}$ to refer to $F_{\mathbf{X}}\left(\mathbf{x}\right)$.  We are interested in the largest input-output mutual information over all configurations of the analog linear combiner. Mathematically, we seek for the capacity $C$, which is defined as
\begin{align}\label{eq:capacity_equation}
    C = \underset{\mathbf{A}}{\max} \;C(\mathbf{A}),
\end{align}
as well as the optimal $\mathbf{A}$ and $F_{\mathbf{X}}$ that achieve this capacity. 

It is known that, under general assumptions, the maximizing input distribution of a channel with finite output cardinality is discrete with finite number of mass points \cite{Dysto:2018}. By combining the concavity of the mutual information over the input distribution and the discreteness of the optimal input distribution, numerical algorithms \cite{Blahut:1972,Huang:2005} can be used to obtain the input distribution that solves problem \eqref{eq:capacity_fixed_A} in the general case. The computation, however, may involve multi-dimensional integration, and the complexity would grow exponentially with the number of sign quantizers. For the special case of $N_{\mathrm{r}} = N_{\mathrm{q}}$ and $\mathbf{A} = \mathbf{I}_{N_{\mathrm{q}}\times N_{\mathrm{q}}}$ (i.e. no analog preprocessing), \cite{Mo:2014} established capacity bounds which are tight in some SNR regimes and under certain channel conditions. 

While $C(\mathbf{A})$ in problem \eqref{eq:capacity_fixed_A} is known to be a concave maximization problem, it is not clear whether $C$ in problem \eqref{eq:capacity_equation} can be solved by some provably optimal method in the quantized setting. This problem formulation was considered in \cite{Khalili:2018}, which showed that $C$ can be attained in the infinite SNR regime by choosing an $\mathbf{A}$ that maximizes the partitions of the transmit signal space. Moreover, $C$ in problem \eqref{eq:capacity_equation} can be upper bounded by the $\mathrm{TSC}=\min\{C_{\mathrm{MIMO-AWGN}},N_{\mathrm{q}}\}$ \cite{Khalili:2020,Khalili:2021}. Here, $C_{\mathrm{MIMO-AWGN}}$ is the MIMO AWGN channel capacity without quantization and $N_{\mathrm{q}}$ is the number of 1-bit quantizers. The TSC bound serves as a \emph{universal} upper bound for the capacity of \emph{any} discrete-time memoryless MIMO channel with \emph{any} $N_{\mathrm{q}}$-bit quantization at the output. This bound, however, does not utilize the information about $\mathbf{t}$. 

In the following section, we present an achievability scheme that frames \eqref{eq:capacity_equation} as a resource allocation problem of the transmit power and sign quantizers over all eigenchannels of $\mathbf{H}$. The resulting transmit power and sign quantizer allocation correspond to  specific choices of $F_{\mathbf{X}}$ and $\mathbf{A}$, respectively, which are not necessarily optimal. Effectively, the rate of this scheme
gives a lower bound on the solution of \eqref{eq:capacity_equation}. We also show that the rate of this achievability scheme is tight in the low SNR regime. However, it does not attain the high SNR capacity when $\min\{N_{\mathrm{t}},N_\mathrm{r}\} < 2N_{\mathrm{q}}$.

\section{Achievability Scheme for the Hybrid One-Shot Receiver}
\label{section-achievability_scheme}

In this section, we establish an achievable result for the capacity of the MIMO system depicted in Figure \ref{fig:mimo_sys_model}. The key idea in this achievability result is to configure $\mathbf{A}$ such that the channel becomes a set of parallel SISO subchannels with phase quantization at the output. We then use the capacity-achieving input for this channel as our transmit strategy. 

First, we apply the singular value decomposition (SVD) to the channel matrix $\mathbf{H}$ to get the following matrix factorization:
\begin{align}\label{eq:H_svd}
    \mathbf{H} = \mathbf{U}\mathbf{\Sigma}\mathbf{V}^{H},
\end{align}
where $\mathbf{\Sigma}\in\mathbb{C}^{N_{\mathrm{r}}\times N_{\mathrm{t}}}$ is a diagonal matrix. The first $N_{\sigma} = \min\left\{N_{\mathrm{t}},N_{\mathrm{r}}\right\}$ diagonal entries are the singular values $\{\sqrt{\lambda_{i}}\}_{i = 1}^{i = N_{\sigma}}$ arranged in non-increasing order (that is $\lambda_{i} \geq \lambda_{i+1}$), and the remaining diagonal entries are zeros. The matrices $\mathbf{U}\in\mathbb{C}^{N_{\mathrm{r}}\times N_{\mathrm{r}}}$ and $\mathbf{V}\in\mathbb{C}^{N_{\mathrm{t}}\times N_{\mathrm{t}}}$ are unitary matrices. Suppose we define $\mathbf{\tilde{x}} = \mathbf{V}\mathbf{x}$ as the precoded transmitted symbols. Then, $\mathbf{y}$ can be written as
\begin{align}\label{eq:sys_model_quantized_achievable1}
    \mathbf{y} =&\; \mathrm{sign}\left(\Re\left\{\mathbf{A}(\mathbf{U}\mathbf{\Sigma}\mathbf{V}^{H}\mathbf{\tilde{x}}) + \mathbf{z}'\right\}\right) \nonumber\\
    =&\; \mathrm{sign}\left(\Re\left\{\mathbf{A}(\mathbf{U}\mathbf{\Sigma}\mathbf{x}) + \mathbf{z}'\right\}\right).
\end{align}
Without loss of generality, we can set the analog linear combiner $\mathbf{A}$ as a product of two matrices $\mathbf{\Phi}\in\mathbb{C}^{N_{\mathrm{q}}\times N_{\mathrm{r}}}$ and $\mathbf{U}^{H}\in\mathbb{C}^{N_{\mathrm{r}}\times N_{\mathrm{r}}}$. Equation (\ref{eq:sys_model_quantized_achievable1}) then simplifies to
\begin{align}
    \mathbf{y} =& \mathrm{sign}\left(\Re\left\{\mathbf{\Phi}(\mathbf{\Sigma}\mathbf{x}) + \mathbf{z}'\right\}\right)\nonumber,
\end{align}
and the optimization problem \eqref{eq:capacity_equation} becomes
\begin{align}\label{eq:achievability}
    C = \underset{\mathbf{\Phi}}{\max}\;\underset{F_\mathbf{X}}{\max} \;I_{\mathbf{\Phi}}(\mathbf{x};\mathbf{y}),
\end{align}
where we used the subscript $\mathbf{\Phi}$ to emphasize the dependence of the mutual information in the choice of $\mathbf{\Phi}$.

\subsection{Design $\mathbf{\Phi}$ to create phase quantizers}

\begin{figure}[t]
    \centering
    \includegraphics[scale = .725]{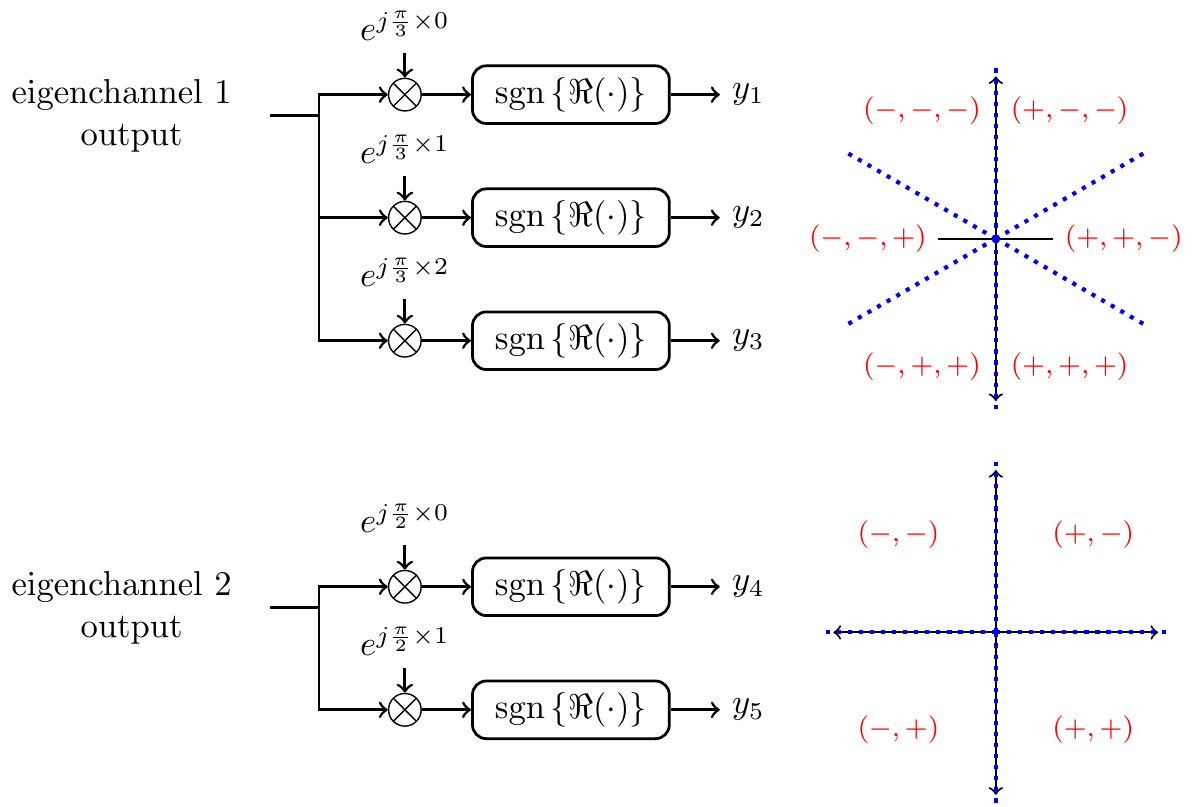}
    \caption{Illustration of how $\mathbf{\Phi}_{\mathrm{PH}}$ is constructed for a MIMO channel with $N_{\sigma} = 2$, $N_{\mathrm{q}} = 5$, and $\mathbf{s} = [3,2]^T$.}
    \label{fig:three_sign_quant}
\end{figure}

Prior to multiplying $\mathbf{\Phi}$, we have $N_{\sigma}$ parallel complex AWGN channels. The matrix $\mathbf{\Phi}$ can be used to create symmetric phase quantizers and connect these phase quantizers to each eigenchannel. First, we define an $N_{\mathrm{r}}\times 1$ quantizer allocation vector $\mathbf{s} = [s_{1},\cdots,s_{N_\sigma},\mathbf{0}_{1\times N_{\mathrm{r}}-N_{\sigma}}]^T$ such that $\sum_{i = 1}^{N_{\sigma}}s_{i} = N_{\mathrm{q}}$ and $s_i \geq 0$. The entries $s_{i}$ correspond to the number of sign comparators used to discretize the output of the $i$-th channel. The set of $s_i$ sign comparators creates a $2s_i$-sector phase quantizer connected at the output of the $i$-th channel. An illustrative example is shown in Figure \ref{fig:three_sign_quant} where three (rotated) sign quantizers are connected at the output of eigenchannel 1 to realize a 6-sector phase-quantized channel and two (rotated) sign quantizers are connected at the output of eigenchannel 2 to realize a 4-sector phase-quantized channel. In this case, we have $\mathbf{s} = [3,2]^T$ and $\mathbf{\Phi}$ becomes
\begin{align*}
    \mathbf{\Phi} = [\boldsymbol{\phi}_{1},\boldsymbol{\phi}_2],\;\;\mathrm{where} \begin{cases}
    \boldsymbol{\phi}_{1} = [1,e^{j\frac{\pi}{3}},e^{j\frac{2\pi}{3}},0,0]^T\\
    \boldsymbol{\phi}_{2} = [0,0,0,1,e^{j\frac{\pi}{2}}]^T
    \end{cases}.
\end{align*}

More generally, the $\mathbf{\Phi}$ matrix is a horizontal stacking of $\boldsymbol{\phi}_k \in\mathbb{C}^{N_{\mathrm{q}}\times 1}$ for  $k = 1,\cdots,N_{\mathrm{r}}$. The first $N_{\sigma}$ column vectors can be expressed as
\begin{align}\label{eq:phi_columns}
    \boldsymbol{\phi}_k = [ \mathbf{0}_{1\times\sum_{i=0}^{k-1}s_{i}}\;\mathbf{e}_{1\times s_k}\;\mathbf{0}_{1\times\sum_{i=k+1}^{N_{\mathrm{q}}}s_{i}}]^T
\end{align}
for all $k\in\{1,\cdots,N_{\sigma}\}$, where the $l$-th entry of $\mathbf{e}_{s_k\times 1}$, denoted as $\mathbf{e}_{s_k\times 1}^{(l)}$, is $e^{\frac{j\pi(l-1)}{s_k}}$. The remaining $N_{\mathrm{r}} - N_{\sigma}$ column vectors of $\mathbf{\Phi}$ are all-zero vectors. Effectively, we define $\mathcal{Q}_{\mathrm{K}}^{\mathrm{PH}}(\cdot): \mathbb{C}\mapsto\mathbb{Z}$ as a function that performs a $K$-sector symmetric phase quantization to map a complex value to an integer value between 0 to $K - 1$. The output of the $i$-th phase-quantized eigenchannel is then
\begin{align}
    y_{\mathrm{q}}^{(i)} = \mathcal{Q}_{2\mathrm{s}_i}^{\mathrm{PH}}\left(\sqrt{\lambda_i}\mathbf{x}^{(i)} + \mathbf{z}^{(i)}\right),
\end{align}
where $\mathbf{x}^{(i)}$ and $\mathbf{z}^{(i)}$ are the $i$-th entry of $\mathbf{x}$ and $\mathbf{z}$, respectively. We shall call this choice of $\mathbf{\Phi}$ as $\mathbf{\Phi}_{\mathrm{PH}}$ and the corresponding analog combiner as $\mathbf{A}_{\mathrm{PH}} = \mathbf{\Phi}_{\mathrm{PH}}\mathbf{U}^H$. Note that
\begin{align}
    C = \underset{\mathbf{A}}{\max} \;C(\mathbf{A}) \geq C(\mathbf{A}_{\mathrm{PH}})
\end{align}
with equality if $\mathbf{A}_{\mathrm{PH}}$ is the analog combiner configuration that maximizes $C(\mathbf{A})$. Thus, the quantity $C(\mathbf{A}_{\mathrm{PH}})$ is the rate of our achievability scheme and is a lower bound for problem \eqref{eq:capacity_equation}.

To solve $C(\mathbf{A}_{\mathrm{PH}})$, we present two function definitions that correspond to the conditional probability and conditional entropy of a Gaussian channel with $K$-sector phase quantization at the output. 

\begin{definition}\label{definition:phase_quantization_func}
The phase quantization probability function, $W_{y}^{(K)}(\nu,\theta)$, is defined as
\begin{equation}\label{eq:phase_quantization_prob}
    W_y^{(K)}(\nu,\theta) = \int_{\frac{2\pi}{K}y-\pi-\theta}^{\frac{2\pi}{K}(y+1)-\pi-\theta}f_{\Phi|N}(\phi|\nu)\;d\phi,
\end{equation}
where $f_{\Phi|N}(\phi|\nu)$ is
\begin{align}\label{eq:prob_phi_given_nu}
   = \frac{e^{-\nu}}{2\pi}+\frac{\sqrt{\nu}\cos\left(\phi\right)e^{-\nu\sin^2\left(\phi\right)}\left[1-Q\left(\sqrt{2\nu}\cos\left(\phi\right)\right)\right]}{\sqrt{\pi}},\end{align}
and $Q(x)$ is the Gaussian Q-function, $\theta\in[-\pi,\pi]$, and $\nu\geq 0$.
\end{definition}

\begin{definition}\label{definition:phase_quantization_entrop}
The phase quantization entropy, $w_{K}(\nu,\theta)$, is defined as\footnote{All $\log()$ terms in this paper are in base 2 unless specified otherwise.}
\begin{align}\label{eq:phase_quantization_entrop}
    w_{K}(\nu,\theta) = -\sum_{y=0}^{K-1}W_y^{(K)}(\nu,\theta)\log W_y^{(K)}(\nu,\theta)
\end{align}
for any $\theta\in[-\pi,\pi]$, $\nu\geq 0$, and $K > 1$. The function $W_y^{(K)}(\nu,\theta)$ is given in Definition \ref{definition:phase_quantization_func}.
\end{definition}

From \cite[Theorem 1]{bernardo2021TIT}, the capacity-achieving input of a Gaussian channel with $K$-sector phase quantization at the output is a rotated $K$-phase shift keying (PSK) and the capacity can be computed numerically. Consequently, we formulate $C(\mathbf{A}_{\mathrm{PH}})$ as
\begingroup
\allowdisplaybreaks
\begin{subequations}\label{eq:achievability_opt}
\begin{alignat}{2}
C(\mathbf{A}_{\mathrm{PH}}) = \max_{s_{\mathrm{i}},\rho_{i}} \quad &   \sum_{i=1}^{N_{\sigma}} C_{\phi}\left(s_{i},\frac{\lambda_{i}\rho_{i}}{\sigma^2}\right)\\
\textrm{s.t.} \quad & \sum_{i = 1}^{N_{\sigma}}s_i = N_{\mathrm{q}}\\
  &\sum_{i = 1}^{N_{\sigma}}\rho_i \leq P\\
  &\qquad\rho_i \geq 0\;, s_{i}\in \{0,1,\cdots,N_{\mathrm{q}}\},
\end{alignat}
\end{subequations}
\endgroup
where $C_{\phi}(s,\nu)$ is the capacity of a scalar Gaussian channel with SNR$=\nu$ and $2s$-sector phase quantization at the output. Mathematically,
\begin{align}\label{eq:C_phi}
    C_{\phi}(s,\nu) = \begin{cases}\log 2 s - w_{2s}\left(\nu,\frac{\pi}{2s}\right),\qquad s \geq 1\\
    \qquad\qquad0\qquad\qquad,\qquad \mathrm{otherwise}
    \end{cases}.
\end{align}

The objective function of \eqref{eq:achievability_opt} is the sum capacity of the $N_{\sigma}$ parallel eigenchannels with phase quantization at the output. A $2s_{i}$-PSK with amplitude $\sqrt{\rho_i}$ is transmitted over the $i$-th eigenchannel to attain the capacity. The optimization of $\{s_i\}_{i = 1}^{N_{\sigma}}$ corresponds to the optimization of $\mathbf{\Phi}_{\mathrm{PH}}$ (consequently, $\mathbf{A}_{\mathrm{PH}}$)  whereas the optimization of $\{\rho_i\}_{i = 1}^{N_{\sigma}}$ gives the optimal $F_{\mathbf{X}}$ that attains C($\mathbf{A}_{\mathrm{PH}}$).

\subsection{Alternating Optimization Approach}
One major issue with optimization problem \eqref{eq:achievability_opt} is its nonconvex stucture due to the discrete parameters in the search space. In this subsection, we present a polynomial-time heuristic method that maximizes the objective function in \eqref{eq:achievability_opt}. The approach is based on alternating optimization of the parameters.

First, we define the parameter $N_{\mathrm{s}}$ to be the number of active\footnote{An eigenchannel is active if it has a nonzero transmit power and a nonzero quantizer allocation. Otherwise, it is inactive.} eigenchannels that we intend to use for transmission. Note that the $N_{\mathrm{s}}$ active eigenchannels in the achievability scheme correspond to the eigenchannels having the $N_{\mathrm{s}}$ strongest singular values. The optimality of this choice is formalized in the following lemma.
\begin{lemma}\label{lemma:active_Ns}
If the optimal strategy has $N_\mathrm{s}$ active eigenchannels, then the eigenvalues of those eigenchannels should be $\{\lambda_i\}_{i = 1}^{i = N_\mathrm{s}}$. In other words, these channels should have the strongest eigenvalues among $N_{\sigma}$ eigenchannels.
\end{lemma}
\begin{proof}
See Appendix \ref{proof_Ns}.
\end{proof}
By Lemma \ref{lemma:active_Ns}, we can rewrite problem \eqref{eq:achievability_opt} without loss of optimality as
\begingroup
\allowdisplaybreaks
\begin{subequations}\label{eq:achievability_opt_v1}
\begin{alignat}{2}
C(\mathbf{A}_{\mathrm{PH}}) = \max_{s_{\mathrm{i}},\rho_{i},N_{\mathrm{s}}} \quad &   \sum_{i=1}^{N_{\mathrm{s}}} \log 2 s_{i} - w_{2s_{i}}\left(\frac{\lambda_{i}\rho_{i}}{\sigma^2},\frac{\pi}{2s_{i}}\right)\\
\textrm{s.t.} \quad & \sum_{i = 1}^{N_{\mathrm{s}}}s_i = N_{\mathrm{q}}\\
  &\sum_{i = 1}^{N_{\mathrm{s}}}\rho_i \leq P\\
  &\qquad \rho_i \geq 0\;, s_{i}\in \{0,\cdots,N_{\mathrm{q}}\}\\
  &\qquad N_{\mathrm{s}}\in\{1,\cdots,N_{\sigma}\}.
\end{alignat}
\end{subequations}
\endgroup
Next, note that for fixed $N_{\mathrm{s}}$ and $\{s_{i}\}_{i = 1}^{N_{\mathrm{s}}}$, the optimization problem \eqref{eq:achievability_opt_v1} can be simplified to
\begingroup
\allowdisplaybreaks
\begin{subequations}\label{eq:achievability_opt_v2}
\begin{alignat}{2}
\min_{ \rho_{i}} \quad &  \sum_{i=1}^{N_{\mathrm{s}}}w_{2s_{i}}\left(\frac{\lambda_{i}\rho_{i}}{\sigma^2},\frac{\pi}{2s_{i}}\right)\\
\textrm{s.t.} \quad &\sum_{i = 1}^{N_{\mathrm{s}}}\rho_i \leq P\\
  &\qquad\rho_i \geq 0\;\forall i\in\{1,2,\cdots,N_{\mathrm{s}}\},
\end{alignat}
\end{subequations}
\endgroup
which has a convex structure. This is because $w_{2s_i}(\nu,\theta)$ is convex on $\nu$ \cite[Proposition 2]{bernardo2021TIT} and the summation of nonnegative convex functions preserves convexity. Consequently, the optimal power allocation, denoted $\{\rho'_{i}\}_{i = 1}^{N_{\mathrm{s}}}$, can be solved efficiently using standard convex solvers.

For fixed $N_{\mathrm{s}}$ and $\{\rho_i\}_{i=1}^{N_{\mathrm{s}}}$, problem \eqref{eq:achievability_opt_v1} becomes a discrete optimization over the parameters $\{s_{i}\}_{i=1}^{N_{\mathrm{s}}}$. This optimal allocation of the sign quantizers, denoted $\{s_{i}'\}_{i = 1}^{N_{\mathrm{s}}}$, can be solved using a dynamic programming approach. We define a state space $\mathcal{S_{\mathrm{state}}}$ with each state being the tuple $(i,n_\mathrm{q})$, where $i \in \{0,\cdots,N_{\mathrm{s}}\}$ and $n_{\mathrm{q}} \in \{0,\cdots,N_{\mathrm{q}}\}$. Define also the function $f(i,n_{\mathrm{q}})$ as the sum capacity of channels 1 to $i$ when there are $n_{\mathrm{q}}$ sign quantizers that can be allocated to these $i$ channels. This function $f(i,n_{\mathrm{q}})$ can be expressed using the recurrence relation in \eqref{eq:DP_quant_alloc}.
\begin{figure*}[b]
\hrulefill
\begin{align}\label{eq:DP_quant_alloc}
    f(i,n_{\mathrm{q}}) = \begin{cases}
    \qquad\qquad\qquad  0 &,\; i = 0\text{ or }n_{\mathrm{q}} = 0\\
      \underset{k\in\{1,\cdots,n_{\mathrm{q}}\}}{\max} \Big\{f(i - 1,n_{\mathrm{q}}-k) + \log 2k - w_{2k}\left(\frac{\lambda_{i} \rho_{i}}{\sigma^2},\frac{\pi}{2k}\right)\Big\}&,\; \mathrm{otherwise}
    \end{cases}.
\end{align}
\end{figure*}
The value of $f(N_{\mathrm{s}},N_{\mathrm{q}})$ gives the sum capacity for $N_{\mathrm{s}}$ active eigenchannels with $N_{\mathrm{q}}$ sign quantizers. We can also compute the optimum choice of $k$ per state $(i,n_{\mathrm{q}})$ as
\begin{align*}
    s(i,n_{\mathrm{q}}) = 
    \underset{k\in\{1,\cdots,n_{\mathrm{q}}\}}{\arg\max} \Bigg\{f(i - 1,n_{\mathrm{q}}-k) + C_{\phi}\left(k,\frac{\lambda_i \rho_i}{\sigma^2}\right)\Bigg\}.
\end{align*}

The proof that the dynamic programming approach solves the optimal quantizer allocation for a fixed $N_{\mathrm{s}}$ and fixed power allocation is presented in Appendix \ref{proof_dp}. We simply iterate over all $N_{\mathrm{s}} \in \{1,\cdots,N_{\sigma}\}$ and then for each value, alternate between the two optimization procedures until convergence. 

\begin{figure*}[t]
    \centering
    \subfloat[]{\label{fig:exact_vs_approx}
    \includegraphics[scale = .6]{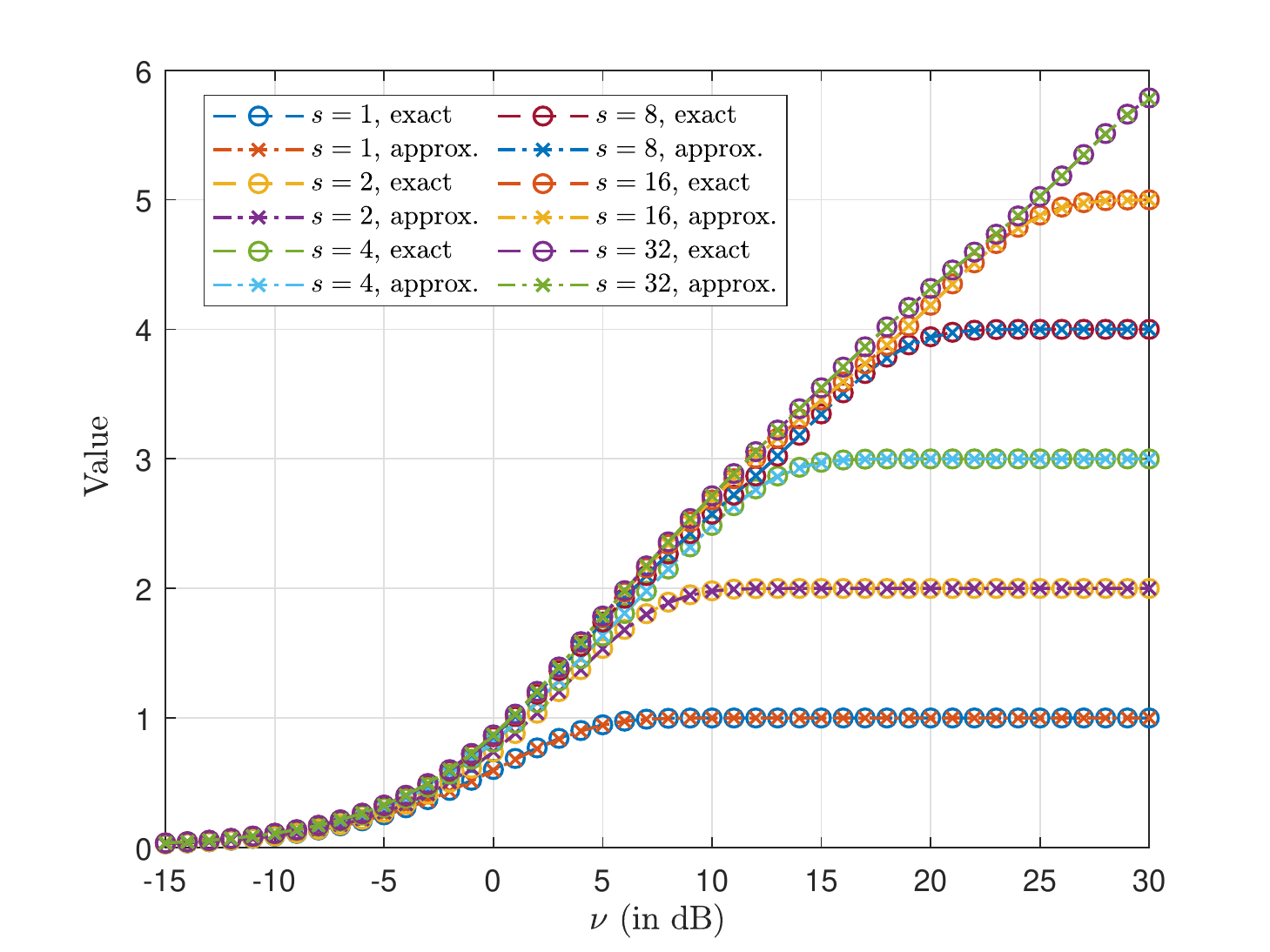}
    }
    \subfloat[]{\label{fig:approx_error}
    \includegraphics[scale = .6]{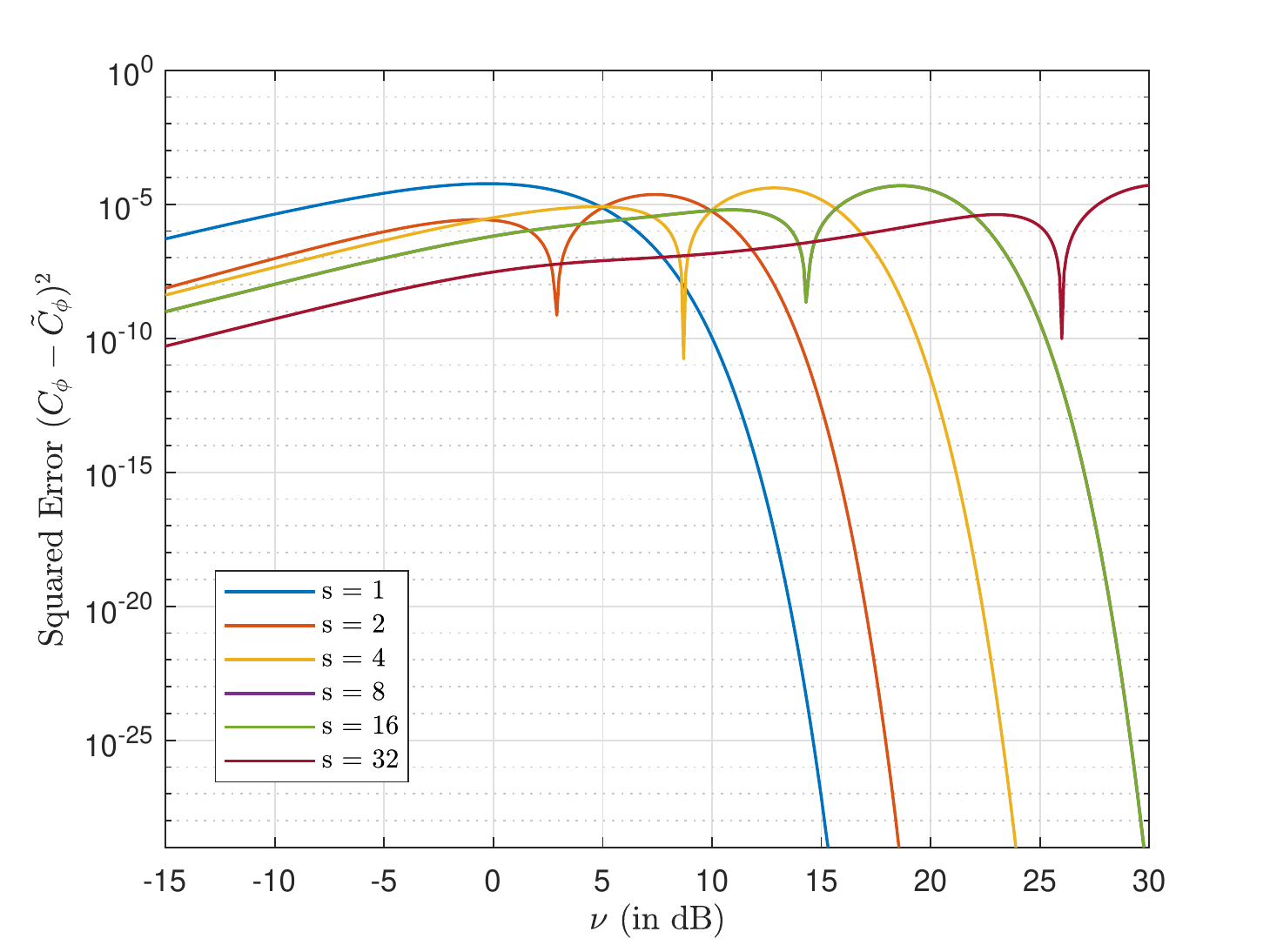}
    }
    
    \caption{(a) Plots of $C_{\phi}(s,\nu)$ and $\tilde{C}_{\phi}(s,\nu)$ for different $s$; and (b) Squared approximation error of $C_{\phi}(s,SNR)$ and $\tilde{C}_{\phi}(s,\nu)$ vs $\nu$ for different values of $s$. We set $R = 9$.}
    
\end{figure*}

The remaining computational bottleneck in solving the optimization problem is the evaluation of $C_{\phi}(s,\nu)$ in \eqref{eq:C_phi}. This is because the phase quantization entropy contains integral terms that need to be computed numerically. To this end, we define a function $\tilde{C}_{\phi}(s,\nu)$ which closely approximates $C_{\phi}(s,\nu)$ as follows:
\begin{align}\label{eq:C_phi_est}
    \tilde{C}_{\phi}(s,\nu) =  \begin{cases}\log 2 s
    -\tilde{w}_{2s}\left(\nu,\frac{\pi}{2s}\right) ,\qquad s \geq 1\\
   \qquad\quad0\quad\qquad\qquad, \qquad \mathrm{otherwise}
    \end{cases},
\end{align}
where
\begin{align}
    \tilde{w}_{2s}(\nu,\frac{\pi}{2s}) = -\sum_{y=0}^{2s-1}\tilde{W}_y^{(2s)}\left(\nu,\frac{\pi}{2s}\right)\log \tilde{W}_y^{(2s)}\left(\nu,\frac{\pi}{2s}\right),
\end{align}
\begin{align}\label{eq:phase_quant_func_est}
    &\tilde{W}_y^{(2s)}\left(\nu,\frac{\pi}{2s}\right) \nonumber\\
    &\qquad= \frac{1}{G}\cdot\frac{\pi}{s R}\sum_{r = 0}^{R-1} f_{\Phi|N}\left(\frac{\pi}{s}\left(y + \frac{r}{R}\right) - \pi -\theta \Big|\nu\right),
\end{align}
and $G = \sum_{y = 0}^{2s-1}\tilde{W}_y^{(2s)}\left(\nu,\frac{\pi}{2s}\right)$. We approximate the integral terms using midpoint rule of definite integrals. $R$ corresponds to the number of rectangles to be used in the approximation and $\frac{\pi}{sR}$ is the width of a rectangle. The factor $1/G$ ensures that the set $\left\{\tilde{W}_y^{(2s)}\left(\nu,\frac{\pi}{2s}\right)\right\}_{y = 0}^{2s-1}$ forms a probability simplex. The plots of $\tilde{C}_{\phi}(s,\nu)$ and $\tilde{C}_{\phi}(s,\nu)$ for different $s$, and the squared approximation error are plotted in Figures \ref{fig:exact_vs_approx} and \ref{fig:approx_error}, respectively. Here, we fix $R = 9$ and observe different values of $s$. At this choice of $R$, we can see small approximation errors for all $\nu$ and $s$ considered. Throughout this paper, the setting $R = 9$ is used for evaluating $\tilde{C}_{\phi}(s,\nu)$.

Algorithm 1 outlines the alternating optimization approach. Here, the power allocation and quantizer allocation steps are referred to as optimization procedure \#1 and \#2, respectively. We use $\tilde{C}_{\phi}(s,\nu)$ instead of $C_{\phi}(s,\nu)$ in these optimization procedures. The optimized $\{\rho_{i}\}_{i = 1}^{N_{\mathrm{s}}}$, $\{s_{i}\}_{i = 1}^{N_{\mathrm{s}}}$, and $N_{\mathrm{s}}$ are then used to construct the approximate solution for the analog linear combiner $\mathbf{A}_{\mathrm{PH}}$ and the distribution $F_{\mathbf{X}}$ that attains $C(\mathbf{A}_{\mathrm{PH}})$. We shall denote these approximate solutions produced by Algorithm 1 as $\hat{\mathbf{A}}_{\mathrm{PH}}$ and $\hat{F}_{\mathbf{X}}$. While \eqref{eq:C_phi_est} is used to optimize $\mathbf{A}_{\mathrm{PH}}$ and $\mathbf{F}_{\mathbf{X}}$, we still use \eqref{eq:C_phi} to evaluate $C^*$ in Line 15. The computational complexity of computing $\hat{\mathbf{A}}_{\mathrm{PH}}$ and $\hat{F}_{\mathbf{X}}$ using Algorithm 1 is
\begin{align}
    O\left(N_{\sigma}\left\{T_{1} +T_{2}\right\}\log\frac{1}{\epsilon_1}\right),
\end{align}
where
\begin{align*}
   T_1 =&\; O\left(\max\left\{N_{\sigma}^{3}T_{\mathrm{cap}}^{3},\mathcal{F}\right\}\log \frac{N_{\sigma}}{\epsilon_2}\right)\\\ T_2 =&\; O\left(N_{\sigma}N_{\mathrm{q}}^2T_{\mathrm{cap}}\right) \\
   T_{\mathrm{cap}} =&\; O(N_{\mathrm{q}}R).
\end{align*}
The quantities $T_1$ and $T_2$ account for the computational complexity of optimization procedures \#1 and \#2, respectively. For optimization procedure \#1, interior point method is used to solve the convex power allocation problem. The expression $\max\left\{N_{\sigma}^{3}T_{\mathrm{cap}}^{3},\mathcal{F}\right\}$ in $T_1$ is the number of operations per iteration \cite[Section 1.3.1]{Boyd:2004_convex-opt} and $\log\left( \frac{N_{\sigma}}{\epsilon_2}\right)$ accounts for the number of iterations \cite[Section 11.3.3]{Boyd:2004_convex-opt}. The parameter $\mathcal{F}$ is the cost of evaluating the first and second derivatives of the objective function and constraints. The parameters $\epsilon_1$ and $\epsilon_2$ set the convergence criterion for the alternating optimization scheme and the optimization procedure \#1, respectively. Finally, $T_{\mathrm{cap}}$ is the computational complexity of evaluating $\tilde{C}_{\phi}(s,\nu)$.

We note that Algorithm 1 only guarantees convergence to a local optimal solution of problem \eqref{eq:achievability_opt}. This is because $C(\mathbf{A}_{\mathrm{PH}})$ is not jointly convex over the parameters $\{\rho_{i}\}$ and $\{s_{i}\}$. Furthermore, $C(\mathbf{A}_{\mathrm{PH}})$ is just a lower bound on $C$. Thus, we have 
\begin{align*}
    C(\hat{\mathbf{A}}_{\mathrm{PH}}) \leq C(\mathbf{A}_{\mathrm{PH}}) \leq C.
\end{align*}
It is therefore necessary to establish an upper bound on \eqref{eq:capacity_equation} to gauge how far, at worst, is $C(\hat{\mathbf{A}}_{\mathrm{PH}})$ to the exact capacity $C$.

\begin{algorithm}[t]\label{algo:alternating_opt}
\DontPrintSemicolon
  \KwInput{$N_{\sigma}$, $\{\lambda_{i}\}_{i = 1}^{N_{\sigma}}$, $\sigma^2$, $N_{\mathrm{q}}$, $P$} 
  \KwOutput{$\hat{\mathbf{A}}_{\mathrm{PH}}$, $\hat{F}_{\mathbf{X}}$, $C^*$ }
  $C^* = 0$, $N_{s}^* = 0$, $N_{\mathrm{s}} = 1$    \tcp*{Initialize}
  Set $\epsilon_1$ \tcp*{Set convergence condition}
  
  
  \For{$N_{\mathrm{s}} = 1\;\mathbf{ to }\;N_{\sigma}$} 
  {
        \tcc{Uniform allocation of $s_{i}$}
        $s_{i} = \big\lceil \frac{N_{\mathrm{q}}}{N_{\mathrm{s}}}\big\rceil\;\forall i\in \{1,\cdots,N_{\mathrm{q}}\;\mathrm{mod}\; N_{\mathrm{s}}\}$,\; $s_{i} = \big\lfloor \frac{N_{\mathrm{q}}}{N_{\mathrm{s}}}\big\rfloor\;\forall i\in \{N_{\mathrm{q}}\;\mathrm{mod}\;  N_{\mathrm{s}}+1,\cdots,N_{\mathrm{s}}\}$
        
        \Repeat{$C_2 - C_1 < \epsilon_1$} {
            
            \tcc{Optimization Procedure \#1 (optimize power allocation)}
            Solve $\{\rho^*_{i}\}_{i = 1}^{N_{\mathrm{s}}}$ using the formulation in (\ref{eq:achievability_opt_v2}).\;
            $C_{1} = \sum_{i=1}^{N_{\mathrm{s}}} \tilde{C}_{\phi}\left(s_i,\frac{\lambda_{i}\rho_i^*}{\sigma^2}\right)$\;
           
            $\rho_{i} = \rho^*_{i} \quad\forall i \in \{1,\cdots,N_{s}\}$
            
            \tcc{Optimization Procedure \#2 (optimize quantizer allocation)}
            Solve $\{s^*_{i}\}_{i = 1}^{N_{\mathrm{s}}}$ using dynamic programming. \;
            $C_{2} = \sum_{i=1}^{N_{\mathrm{s}}}\tilde{C}_{\phi}\left(s_i^*,\frac{\lambda_i\rho_i}{\sigma^2}\right)$ \;
            $s_{i} = s^*_{i} \quad\forall i \in \{1,\cdots,N_{s}\}$\;
        }
        \If{$C_2 > C^*$ and $s_i > 0\;\forall i\in\{1,\cdots,N_{s}\}$ \tcp*{Update optimal values}} {
            $C^* = \sum_{i=1}^{N_{\mathrm{s}}}\log 2 s_{i} - w_{2s_{i}}\left(\frac{\lambda_{i}\rho_{i}}{\sigma^2},\frac{\pi}{2s_{i}}\right)$\;
            $N^*_{\mathrm{s}} = N_{\mathrm{s}}$,\quad $s'_{i} = s_{i}$, $\rho'_{i} = \rho_{i} \quad\forall i \in \{1,\cdots,N^*_{s}\}$\;
        }
  }
  $\mathbf{\Phi}_{\mathrm{PH}} = [\boldsymbol{\phi}_{1},\cdots,\boldsymbol{\phi}_{N_{\mathrm{s}}^*},\cdots, \mathbf{0}]$, where $\boldsymbol{\phi}_k$ is from \eqref{eq:phi_columns}.\; 
  $\hat{\mathbf{A}}_{\mathrm{PH}} = \mathbf{\Phi}_{\mathrm{PH}}\mathbf{U}^H$\;
  $\hat{F}_{\mathbf{X}_i} = 2s_i'$-PSK with amplitude $\sqrt{\rho_i'}$ 
\caption{Alternating Optimization to get local optimal solution to Problem (\ref{eq:achievability_opt})}
\end{algorithm}

\section{Capacity Upper Bound for the Hybrid One-Shot Receiver}
\label{section-cap_upperbound}

\subsection{Capacity Upper Bound 1}\label{subsection-cap_upperbound_Inf}

We now derive an upper bound for the capacity expression in \eqref{eq:capacity_equation}. First, we use the high SNR capacity results in \cite{Khalili:2018} and apply it to $\mathbb{C}^{N_{\sigma}}$ to establish $C$ in the infinite SNR regime, which we denote as $C_{\infty}$. That is, we maximize the number of regions created by arranging the hyperplanes in $N_{\sigma}$ complex dimensions. By doing this, we maximize the output entropy. For the special case where all the hyperplanes intersect the origin, \cite{Ho:2006} proved that the number of regions created by $N_{\mathrm{q}}$ hyperplanes in $2N_{\sigma}$ real dimensions is
\begin{align}
    \mathcal{M}_{\infty} = 2\sum_{i = 0}^{2N_{\sigma}-1}\binom{N_{\mathrm{q}}-1}{i},
\end{align}
where $\binom{n}{m} = 0$ if $m > n$. Consequently, $C_{\infty} = \log \mathcal{M}_{\infty}$ by choosing an $F_{\mathbf{X}}$ that induces a uniform output distribution. 

\subsection{Capacity Upper Bound 2}
\label{subsection-cap_upperbound_finite}

\begin{figure}[t]
    \centering
    \includegraphics[scale = .68]{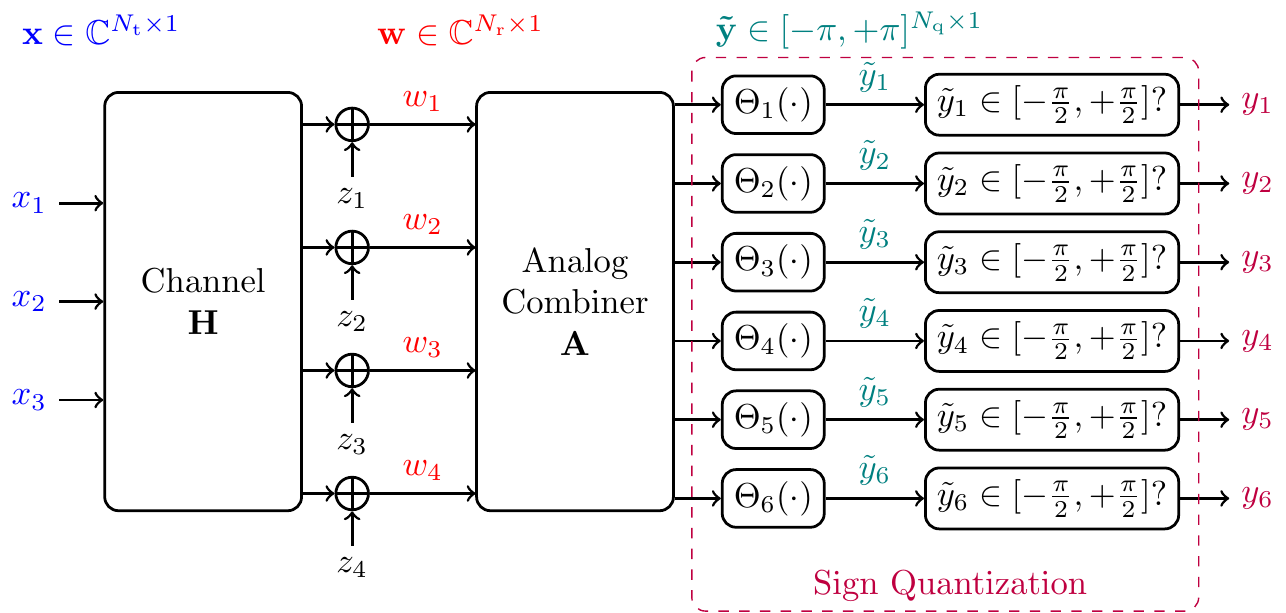}
    \caption{Modified System Model with Sign Quantization broken down to two-stages: (1) Phase Detection and (2) Phase-to-bit.}
    \label{fig:mimo_sys_model_converse}
\end{figure}

The capacity upper bound established in the previous subsection is quite loose in the finite SNR case. To establish a tighter capacity upper bound in the finite SNR regime, we note that the sign quantizer only requires the phase information of its input in order to output $-1$ or $+1$. Suppose we define a function $\Theta_{k}(\cdot)$ that extracts the phase information of the $k$-th output of the analog combiner, and another function that maps the phase detector output to $+1$ when it is inside the region $[-\frac{\pi}{2},\frac{\pi}{2}]$ and is mapped to $-1$ otherwise. Then, the system model in Figure \ref{fig:mimo_sys_model_converse} is the same as the system model depicted in Figure \ref{fig:mimo_sys_model}. It then follows that the capacity of the two system models are equal. By the DPI, we have
\begin{align*}
    I_{\mathbf{A}}(\mathbf{x};\mathbf{y}) \leq I_{\mathbf{A}}(\mathbf{x};\tilde{\mathbf{y}}),
\end{align*}
and an upper bound on the channel capacity can be established by maximizing the mutual information between the input $\mathbf{x}$ and the outputs of the phase detectors. Define $\mathbf{\Theta}_{i:j}$ to be the phase detector outputs from index $i$ to index $j$, and $\mathbf{z}'_{i:j}$ to be the noise components at the output of the analog combiner from index $i$ to index $j$. The following lemma further simplifies the capacity maximization problem. 

\begin{lemma}\label{lemma:reduce_branches}
Suppose we have $N_{\sigma} \leq N_{\mathrm{q}}$. Suppose further that there exists an analog linear combiner $\mathbf{A} = \begin{bmatrix}\mathbf{A}_1\\
\mathbf{A}_2\end{bmatrix}$ such that $\mathbf{AH} = \begin{bmatrix}\mathbf{A}_1\mathbf{H}\\
\mathbf{A}_2\mathbf{H}\end{bmatrix} = \begin{bmatrix}\mathbf{B}_1\\
\mathbf{B}_2\end{bmatrix}$, where $\mathbf{B}_1\in\mathbb{C}^{N_{\sigma}\times N_{\mathrm{t}}}$ and $\mathbf{B}_2\in\mathbb{C}^{(N_{\mathrm{q}}-N_{\sigma})\times N_{\mathrm{t}}}$. If $\mathbf{B}_1$ is full rank and $\tilde{\mathbf{y}} = [\tilde{\mathbf{y}}^{(1)}\;\tilde{\mathbf{y}}^{(2)}]^H$, where $\tilde{\mathbf{y}}^{(1)} = \mathbf{\Theta}_{1:N_{\sigma}}(\mathbf{B}_1\mathbf{x} + \mathbf{z}'_{1:N_{\sigma}})$ and $\tilde{\mathbf{y}}^{(2)} = \mathbf{\Theta}_{N_{\sigma}+1:N_{\mathrm{q}}}(\mathbf{B}_2\mathbf{x} + \mathbf{z}'_{N_{\sigma}+1:N_{\mathrm{q}}})$, then
\begin{align*}
    I_{\mathbf{A}}(\mathbf{x};\tilde{\mathbf{y}}) = I_{\mathbf{A}_1}(\mathbf{x};\tilde{\mathbf{y}}^{(1)}).
\end{align*}
Moreover, if $I_{\mathbf{A}}(\mathbf{x};\tilde{\mathbf{y}})$ is achieved by a $\mathbf{B}_1$ that is not full rank, then there exists an $\mathbf{A}'\mathbf{H} = \begin{bmatrix}\mathbf{B}_1'\\\mathbf{B}_2'\end{bmatrix}$ (where $\mathbf{B}_1'$ being full rank) and a distribution $F_{\mathbf{X}'}$ such that $I_{\mathbf{A}}(\mathbf{x};\tilde{\mathbf{y}}) = I_{\mathbf{A}'}(\mathbf{x}';\tilde{\mathbf{y}})$.
\end{lemma}

\begin{proof}
See Appendix \ref{proof_reduce_branches}.
\end{proof}
To put it simply, Lemma \ref{lemma:reduce_branches} shows that we only need to consider $N_{\sigma}$ phase detector outputs since considering more phase detector outputs than $N_{\sigma}$ does not increase the mutual information. Thus, without loss of generality, we can consider the maximization of $I_{\mathbf{A}_1}(\mathbf{x};\tilde{\mathbf{y}}^{(1)})$ with $\mathbf{B}_1 = \mathbf{A}_1\mathbf{H}$ being full rank. In the remainder, we will show that $C$ in problem \eqref{eq:capacity_equation} can be upper bounded by the capacity of a Gaussian MIMO channel with phase detectors at the output. The capacity-achieving input of this channel is also characterized.

Suppose we denote $\tilde{y}_{i}^{(1)}$ as the $i$-th element of $\tilde{\mathbf{y}}^{(1)}$, $\mathbf{\tilde{y}}_{i:j}^{(1)}$ as the elements of $\tilde{\mathbf{y}}^{(1)}$ from index $i$ to index $j$, and $h_{\mathbf{A}_1}(\cdot)$ as the differential entropy function induced by $\mathbf{A}_{1}$. Then, we get the following upper bound on $\max_{\mathbf{A}_1}\max_{F_{\mathbf{X}}}I_{\mathbf{A}_1}(\mathbf{x};\tilde{\mathbf{y}}^{(1)})$:
\begingroup
\allowdisplaybreaks
\begin{align*}
   =& \max_{\mathbf{A}_1}\;\max_{F_{\mathbf{X}}}\;h_{\mathbf{A}_{1}}(\tilde{\mathbf{y}}^{(1)}) - h_{\mathbf{A}_1}(\tilde{\mathbf{y}}^{(1)},\mathbf{x})\\
    =& \max_{\mathbf{A}_1}\;\max_{F_{\mathbf{X}}}\;\sum_{i = 1}^{N_{\sigma}}h_{\mathbf{A}_1}(\tilde{y}^{(1)}_i|\mathbf{\tilde{y}}_{1: i-1}^{(1)}) - \sum_{i = 1}^{N_{\sigma}}h_{\mathbf{A}_1}(\tilde{y}^{(1)}_i|\mathbf{\tilde{y}}_{1: i-1}^{(1)},\mathbf{x})\\
    \overset{(a)}{\leq}& \max_{\mathbf{A}_1}\;\max_{F_{\mathbf{X}}}\; \sum_{i = 1}^{N_{\sigma}}h_{\mathbf{A}_1}(\tilde{y}^{(1)}_i) - \sum_{i = 1}^{N_{\sigma}}h_{\mathbf{A}_1}(\tilde{y}^{(1)}_i|\mathbf{\tilde{y}}_{1: i-1}^{(1)},\mathbf{x})\\
     \overset{(b)}{\leq}& \max_{\mathbf{A}_1}\;\max_{F_{\mathbf{X}}} \sum_{i = 1}^{N_{\sigma}}h_{\mathbf{A}_1}(\tilde{y}^{(1)}_i) - \sum_{i = 1}^{N_{\sigma}}h_{\mathbf{A}_1}(\tilde{y}^{(1)}_i|\mathbf{\tilde{y}}_{1: i-1}^{(1)},\mathbf{\tilde{y}}_{i+1: N_{\sigma}}^{(1)},\mathbf{x})\\
     =& \max_{\tilde{\mathbf{A}}_1\;\mathrm{s.t.}\;\mathbf{B}_1\text{has orthogonal rows}}\max_{F_{\mathbf{X}}}\sum_{i = 1}^{N_{\sigma}}\left[h_{\tilde{\mathbf{A}}_1}(\tilde{y}^{(1)}_i) - h_{\tilde{\mathbf{A}}_1}(\tilde{y}^{(1)}_i|\mathbf{x})\right]\\
     =&  \max_{\tilde{\mathbf{A}}_1\;\mathrm{s.t.}\;\mathbf{B}_1\text{has orthogonal rows}}\;\max_{F_{\mathbf{X}}}\;\sum_{i = 1}^{N_{\sigma}}I_{\tilde{\mathbf{A}}_1}(\mathbf{x};\tilde{y}^{(1)}_i)\\
     =& \max_{F_{\mathbf{X}}}\;\sum_{i = 1}^{N_{\sigma}}I_{\mathbf{U}_1^H}(\mathbf{x};\tilde{y}^{(1)}_i).
\end{align*}
\endgroup
\begin{figure*}[b]
\hrulefill
\begin{align}\label{eq:capacity_UB}
    C_{\phi-\mathrm{detector}} = N_{\sigma}\log(2\pi) - \min_{\rho_i:\sum_{i}\rho_i = P} \left\{\sum_{i = 1}^{N_{\sigma}} \int_{-\pi}^{\pi} f_{\Phi|N}\left(\phi\Big|\frac{\lambda_{i}\rho_{i}}{\sigma^2}\right)\log \frac{1}{f_{\Phi|N}\left(\phi\Big|\frac{\lambda_{i}\rho_{i}}{\sigma^2}\right)}\;d\phi\right\},
\end{align}
where $f_{\Phi|N}\left(\phi|\nu\right)$ is given in \eqref{eq:prob_phi_given_nu}.
\end{figure*}
The first two lines follow from the definition of mutual information and the use of the chain rule. The inequalities (a) and (b) follow from the fact that conditioning reduces entropy. Note that equality in both (a) and (b) can be achieved if and only if $\tilde{y}^{(1)}_i$ is independent of $\mathbf{\tilde{y}}_{1: i-1}^{(1)}$ and $\mathbf{\tilde{y}}_{i+1: N_{\sigma}}^{(1)}$. Since we are able to choose $\mathbf{A}_1$, an appropriate choice of $\mathbf{A}_1$ that achieves equality in the third and fourth line should make $\mathbf{B}_1$ have mutually orthogonal rows. Existence of such $\mathbf{B}_1$ is guaranteed since $N_{\sigma} \leq N_{\mathrm{t}}$. Doing so gives us the fifth line. We call this choice $\tilde{\mathbf{A}}_1$. The sixth line follows from the definition of mutual information. Finally, we obtain the last line using the fact that, out of all the orthonormal bases for the range of $\mathbf{H}$, SVD produces the orthonormal basis for which the total channel gain along each direction is maximized\footnote{Recall the Maximum Variance Formulation of Principal Component Analysis (PCA) and its connection to SVD.}. Thus, $\tilde{\mathbf{A}}_{1}$ should be $\mathbf{U}_{1}^H$, where $\mathbf{U}_{1}$ contains the columns of $\mathbf{U}$ corresponding to the $N_{\sigma}$ nonzero eigenvalues. Consequently, $\mathbf{B}_1 = \Sigma_{N_{\sigma}}$ (the first $N_{\sigma}$ rows of $\Sigma$).

The above result shows that $I_{\mathbf{A}_1}(\mathbf{x};\tilde{\mathbf{y}}^{(1)})$ can be upper bounded by the input-output mutual information of a Gaussian product channel with $N_{\sigma}$ eigenchannels; each eigenchannel has a phase detector at the output. Using this, we can now establish the capacity upper bound in the finite SNR.

\begin{proposition}\label{proposition:finite_SNR_upperbound}
The solution to \eqref{eq:capacity_equation} can be upper bounded by \eqref{eq:capacity_UB}.
\end{proposition}

\begin{proof}
By considering the limiting case of \cite[Theorem 1]{bernardo2021TIT} with $b \rightarrow \infty$, the capacity-achieving input of the $i$-th Gaussian channel with phase detector at the output is an $\infty$-PSK (i.e. a circle with radius $\sqrt{\rho_i}$) for all SNR. Optimal power allocation is applied to the eigenchannels to get the desired capacity upper bound result.
\end{proof}
The first and second terms of \eqref{eq:capacity_UB} are the differential output entropy and conditional differential entropy, respectively. Problem \eqref{eq:capacity_UB} is a convex optimization problem. The convexity of the objective function in \eqref{eq:capacity_UB} comes from \cite[Proposition 2]{bernardo2021TIT}. Finally, by combining \eqref{eq:capacity_UB} and $C_{\infty}$, we get the upper bound
\begin{align}\label{eq:C_upperbound}
    C_{\mathrm{UB}} = \min\left\{C_{\phi-\mathrm{detector}},C_{\infty}\right\}.
\end{align}
For the zero threshold ADC case, this capacity upper bound is tighter than the TSC bound established in \cite{Khalili:2021} since it takes into account the fact that sign quantization throws away the amplitude information. Currently, the analysis framework we developed only applies to the $\mathbf{t} = \mathbf{0}$ case. Extension of the analysis framework to general $\mathbf{t}$ is a potential subject for future work and we are exploring proof techniques that can be used to extend this current framework to general $\mathbf{t}$.  In the next section, we compare \eqref{eq:C_upperbound} and the TSC bound.

\section{Numerical Evaluation of the Achievability Scheme and Upper Bound}
\label{section-numerical_result1}

\begin{figure*}[t]
  \centering
  \subfloat[]{
\includegraphics[width =.525\textwidth]{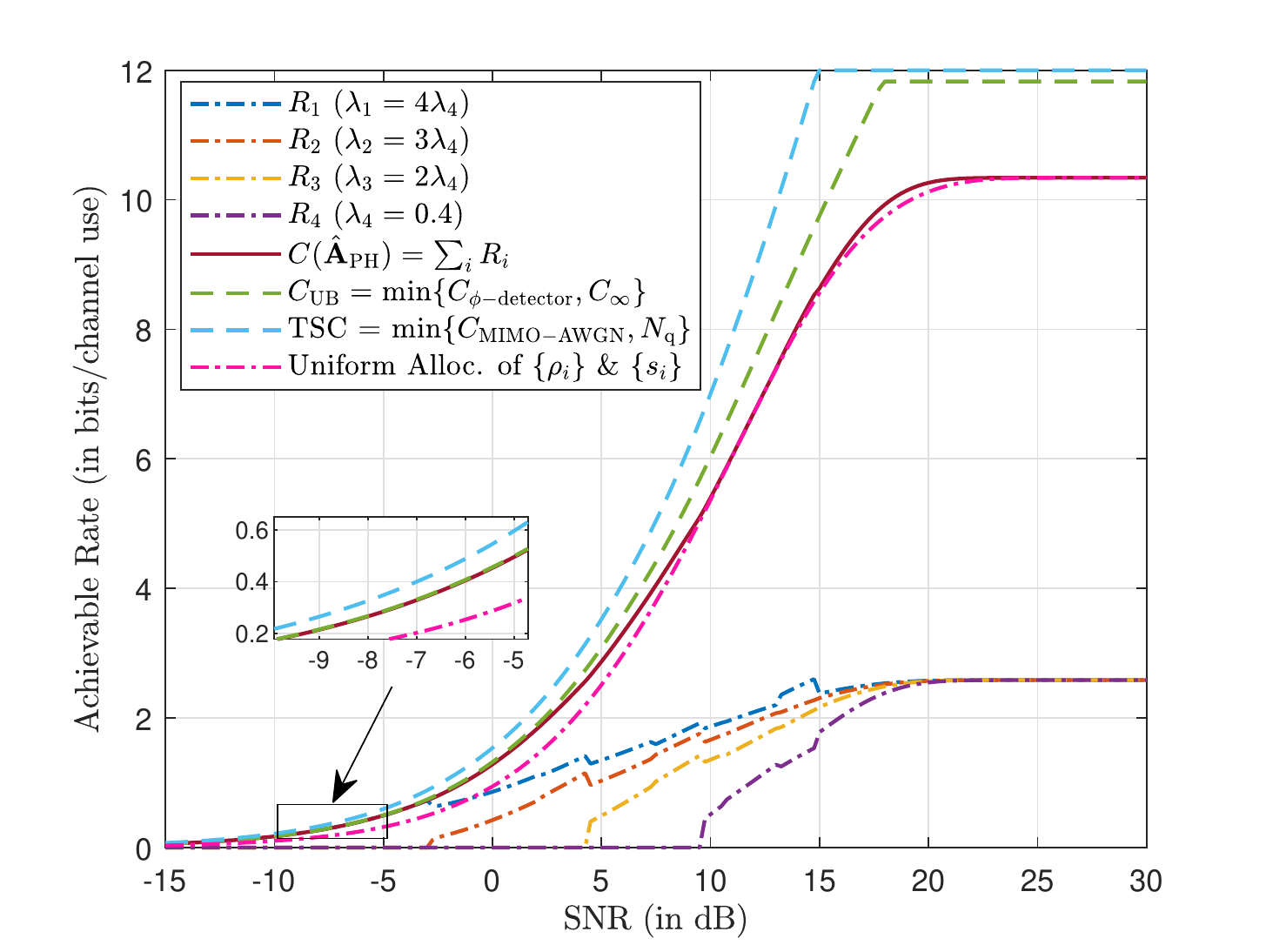}
\label{fig:rate_vs_snr}
}
\hspace*{-.75cm}%
 \subfloat[]{
\includegraphics[width =0.525\textwidth]{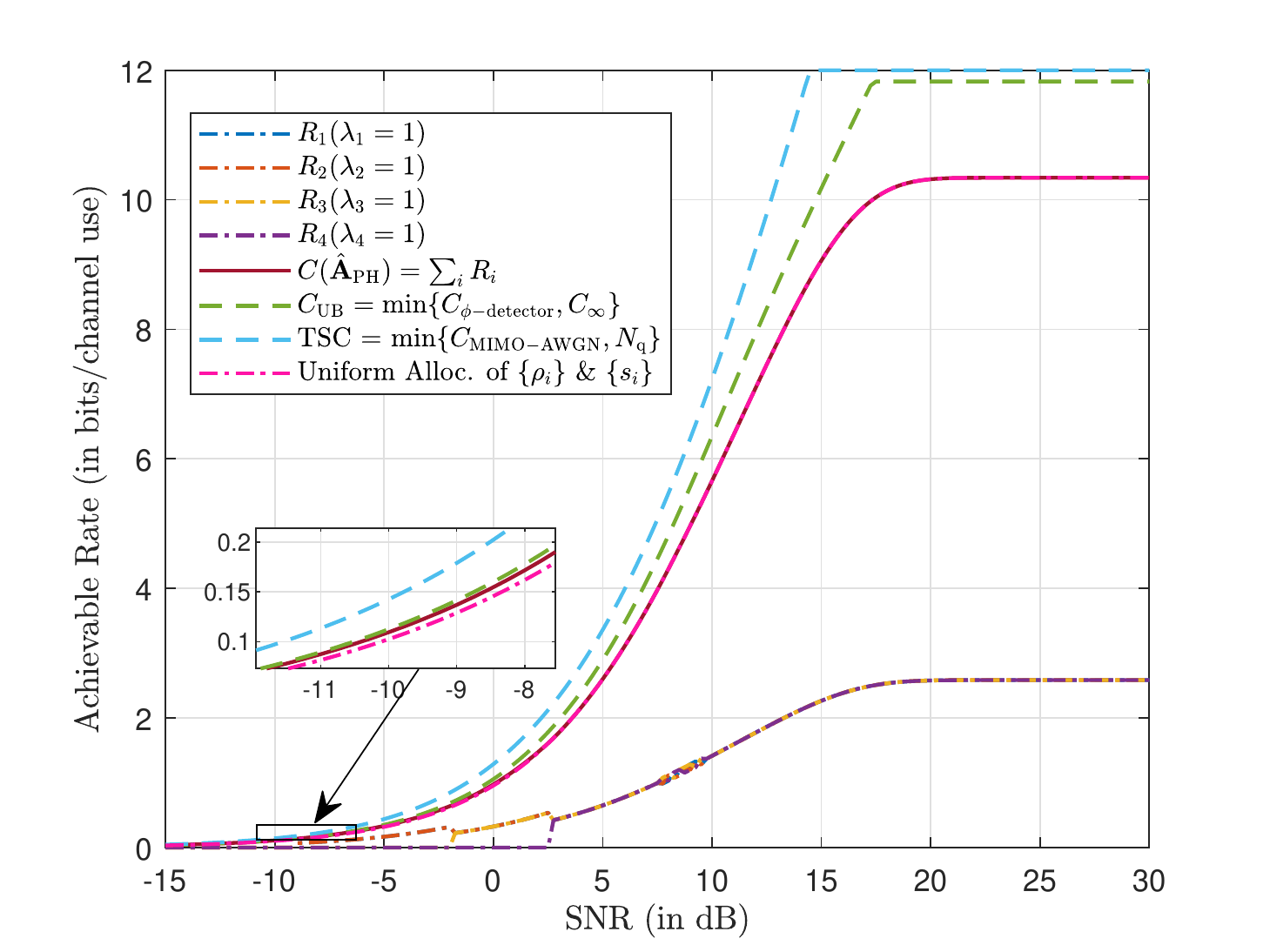}
    \label{fig:rate_vs_snr_equalLambda}
} 
     \caption{Rate vs. SNR of the Achievability Scheme when (a) $\{\lambda_{i}\}_{i = 1}^{4} = \{1.6,1.2,0.8,0.4\}$ and when (b) $\{\lambda_{i}\}_{i = 1}^{4} = \{1.0,1.0,1.0,1.0\}$. Also superimposed are the individual rates, TSC bound, proposed upper bound, and the achievable rate under uniform allocation of $\{\rho_i\}$ and $\{s_i\}$.}
\end{figure*}

In this section, we investigate the performance of our achievability scheme and how close it is to our established capacity upper bound in \eqref{eq:C_upperbound}. We consider a fixed 4$\times$4 channel $\mathbf{H}$ with eigenvalues $\lambda_1 = 1.6, \lambda_2 = 1.2, \lambda_3 = 0.8,$ and $\lambda_4 = 0.4$. We also set $P = 1$ and vary the SNR by changing the noise variance $\sigma^2$. We fix the number of sign quantizers to $N_{\mathrm{q}} = 12$. Even though a 4$\times$4 MIMO setup is considered to generate the numerical results, we note that the insights obtained in this small setup are applicable to larger MIMO settings.

\begin{figure*}[t]
  \centering
  \subfloat[]{
\includegraphics[width =.525\textwidth]{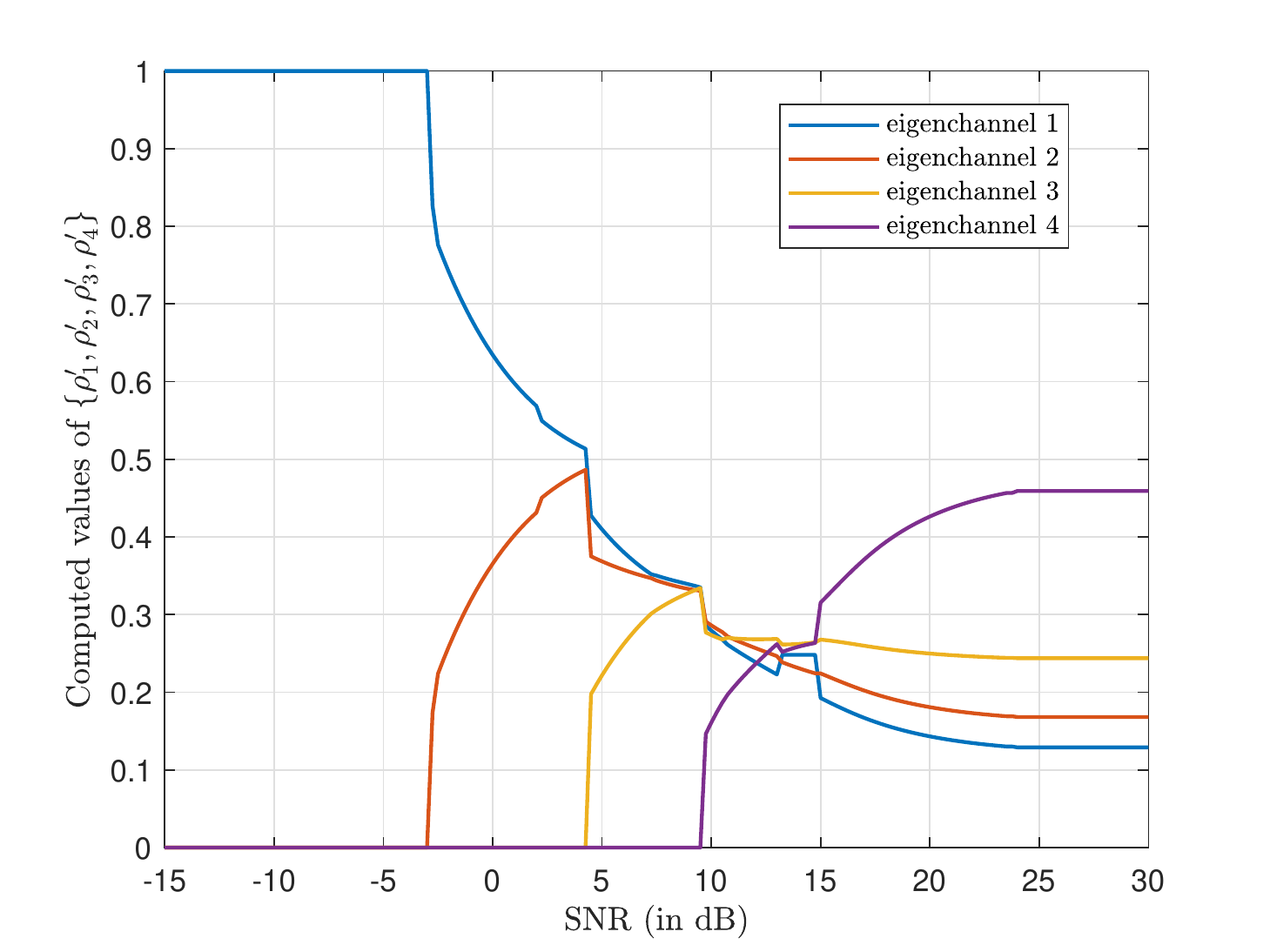}
\label{fig:opt_power}
}
\hspace*{-.75cm}%
 \subfloat[]{
\includegraphics[width =0.525\textwidth]{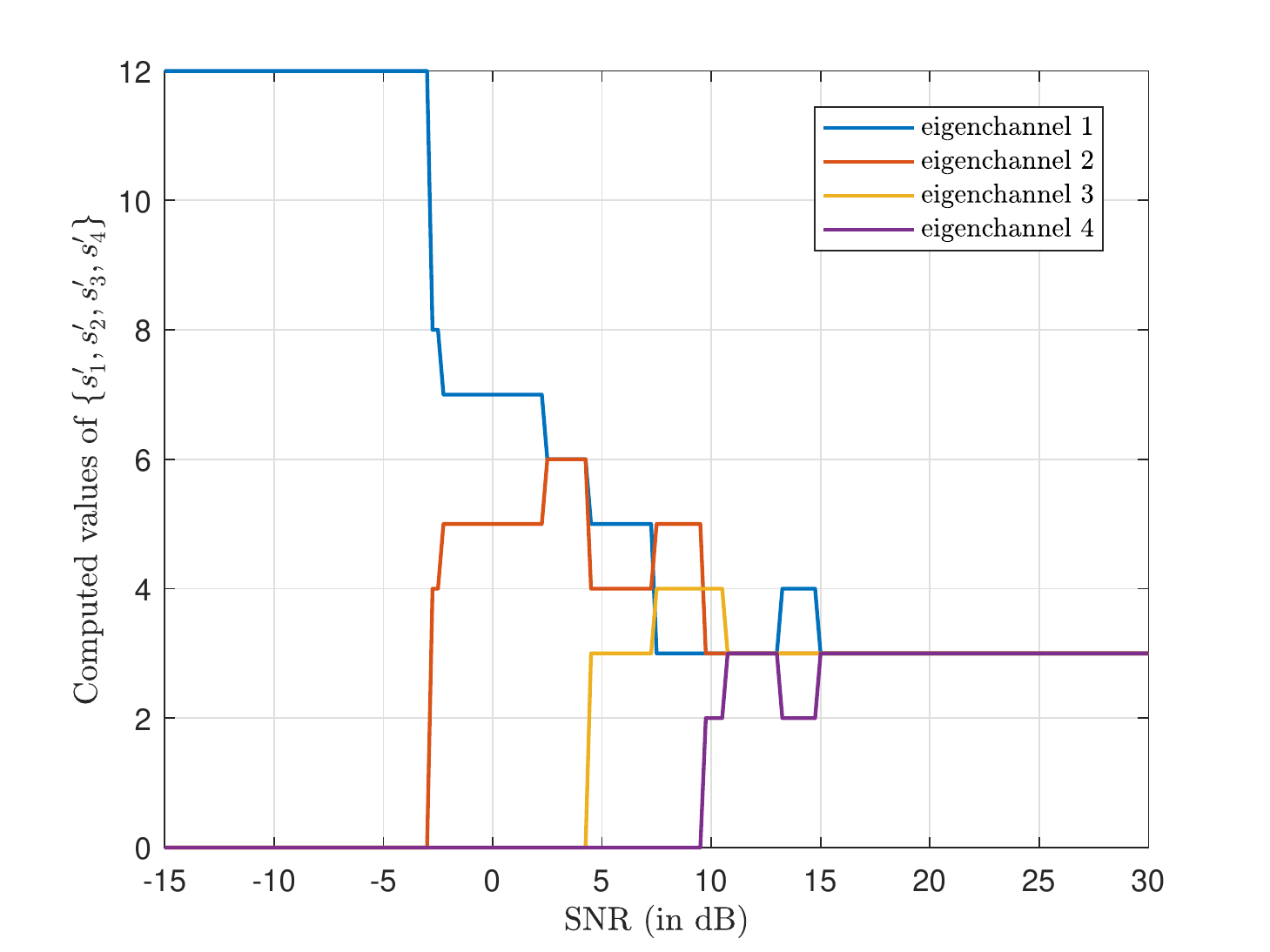}
    \label{fig:opt_quant}
} 
     \caption{Computed values of (a) $\{\rho_i'\}_{i=1}^{4}$ and (b) $\{s_i'\}_{i=1}^{4}$  as a function of SNR for Figure \ref{fig:rate_vs_snr}.}
     \label{fig:opt_allocation}
\end{figure*}

The achievable rate of Algorithm 1, denoted $C(\hat{\mathbf{A}}_{\mathrm{PH}})$, is depicted in Figure \ref{fig:rate_vs_snr}. The individual rates of each eigenchannel are also given in Figure \ref{fig:rate_vs_snr} to see how the rate at each eigenchannel changes with SNR. For comparison, we also superimpose the capacity upper bound given in \eqref{eq:C_upperbound} and the TSC bound. Note that the gap between the TSC bound and our capacity upper bound corresponds to the rate loss incurred when the amplitude information of the received signal is thrown away in a MIMO Gaussian channel. The values of $\{\rho_i'\}_{i = 1}^{N_{\mathrm{s}}}$ and $\{s_i'\}_{i = 1}^{N_{\mathrm{s}}}$ computed by Algorithm 1 are shown in Figures \ref{fig:opt_power} and \ref{fig:opt_quant}, respectively. These parameters are used to construct $\hat{\mathbf{A}}_{\mathrm{PH}}$ and $\hat{F}_{\mathbf{X}}$ according to Lines 17-19 of Algorithm 1. 

It can be observed that $C(\hat{\mathbf{A}}_{\mathrm{PH}})$ is tight in the low SNR regime and outperforms the naive approach of simply setting the $\rho_i$'s and $s_i$'s to be equal. In this regime, the optimal strategy is for the transmitter to use $2N_{\mathrm{q}}$-PSK signaling and send the symbols over the strongest eigenchannel. Simultaneously, the receiver configures $\mathbf{A}_{\mathrm{PH}}$ in such a way that all sign quantizers are connected to the output of the strongest eigenchannel. With $N_{\mathrm{q}} \rightarrow \infty$, the transmission strategy produced by the achievability scheme is an $\infty$-PSK sent over the strongest eigenchannel. Meanwhile, the analog linear combiner is configured to form an $\infty$-bit phase quantizer (or a phase detector). Thus, the gap between the achievable rate and capacity upper bound in the low SNR regime vanishes as $N_{\mathrm{q}}$ grows unbounded. There also exists SNR thresholds, above which we activate the strongest inactive eigenchannel for transmission. The individual rates in Figure \ref{fig:rate_vs_snr} have nonmonotonic behavior and sharp transitions within the SNR range considered. This can be attributed to the discrete nature of optimizing $\{s_i\}$. Nonetheless, the achievable rate remains smooth. 


At this point, one might expect that uniform allocation would have the same performance as Algorithm 1 in all SNR regimes if $\mathbf{H} = \mathbf{I}_{N_{\sigma}\times N_{\sigma}}$ (i.e. eigenvalues are equal). In the classical waterfilling scheme for Gaussian channels, there is no loss of optimality if power is uniformly allocated among subchannels with identical eigenvalues. However, as depicted in Figure \ref{fig:rate_vs_snr_equalLambda}, there is a gap between $C(\mathbf{A}_{\mathrm{PH}})$ and the achievable rate of the hybrid one-shot receiver under uniform allocation of $\{\rho_i\}$ and $\{s_i\}$ in the low SNR regime. Thus, there is still some benefit, albeit small, in optimizing $\{\rho_i\}$ and $\{s_i\}$ when $\mathbf{H} = \mathbf{I}_{N_{\sigma}\times N_{\sigma}}$.

Another intriguing observation in our numerical results is that the values of $\{\rho_i'\}_{i=1}^{N_{\mathrm{s}}}$ and $\{s_i'\}_{i=1}^{N_{\mathrm{s}}}$ given in Figures \ref{fig:opt_power}  and \ref{fig:opt_quant} do not always favor the eigenchannel with the largest eigenvalue. For instance, the strongest eigenchannel does not get the largest share in the available quantizers at SNR = 9 dB and SNR = 10 dB. To validate the result of Algorithm 1, we perform exhaustive search on the optimal $\{\rho_i\}_{i=1}^{N_{\sigma}}$ and $\{s_i\}_{i=1}^{N_{\sigma}}$ to get $C(\mathbf{A}_{\mathrm{PH}})$. That is, we solve the convex power allocation strategy in (\ref{eq:achievability_opt_v2}) for all possible configurations of $\{s_i\}_{i=1}^{N_{\sigma}}$. This guarantees finding the global optimal solution in problem \eqref{eq:achievability_opt}. There are a total of $\binom{N_{\mathrm{q}} + N_{\sigma} - 1}{N_{\sigma} - 1}$ configurations that satisfy $\sum_{i = 1}^{N_{\sigma}}s_{i} = N_{\mathrm{q}}$. We tabulate the top 5 solutions for the joint optimization of $\{\rho_i\}_{i=1}^{N_{\sigma}}$ and $\{s_i\}_{i=1}^{N_{\sigma}}$ for SNR = 9 dB and SNR  = 10 dB in Tables \ref{tab:exhaustive_9dB} and \ref{tab:exhaustive_10dB}, respectively. It can be seen that the optimal $\{\rho_i\}_{i = 1}^{N_{\sigma}}$ and $\{s_i\}_{i = 1}^{N_{\sigma}}$ (marked in blue) obtained by exhaustive search match those produced by Algorithm 1.

\begin{table*}[t]
\renewcommand{\arraystretch}{1}
\caption{Top 5 (out of 455) joint optimization of $\{\rho_i\}_{i = 1}^{N_{\sigma}}$ and $\{s_i\}_{i = 1}^{N_{\sigma}}$ produced by exhaustive search for (a) SNR = 9 dB and (b) SNR = 10 dB. Here, $N_{\sigma} = 4$, $N_{\mathrm{q}} = 12$, and $\{\lambda_i\}_{i = 1}^{4} = \{1.6,1.2,0.8,0.4\}$.}

\centering
\subfloat[][]{\label{tab:exhaustive_9dB}
\begin{tabular}{cccccccccc}
\hline
\textbf{Rank \#} & $R$ (in bpcu) & $s_{1}$ & $s_{2}$ & $s_{3}$ & $s_{4}$ & $\rho_{1}$ & $\rho_{2}$ & $\rho_{3}$ & $\rho_{4}$  \\\hline\hline
\cellcolor{blue!25}1 & \cellcolor{blue!25}4.8349 & \cellcolor{blue!25}3   & \cellcolor{blue!25}5   &  \cellcolor{blue!25}4  & \cellcolor{blue!25}0   & \cellcolor{blue!25}0.3388 & \cellcolor{blue!25}0.3328 &  \cellcolor{blue!25}0.3284   & \cellcolor{blue!25}1.0614e-32\\\hline
2   & 4.8299   &  3   &   4   &   5   &   0  &   0.33622   & 0.32926  &  0.33452  &  2.3924e-32\\\hline
3  &  4.8245   &  5    &  3  &    4    &  0  &   0.32269   & 0.34696  &  0.33035       &      0\\\hline
4  &  4.8237   &  3   &   6   &   3  &    0  &   0.33583  &  0.33724  &  0.32692 &   1.4565e-33\\\hline
 5 &   4.8217   &  4     & 3    &  5  &    0  &   0.32537  &  0.34113  &   0.3335   &  1.254e-32\\\hline
\end{tabular}}%
\quad
\subfloat[][]{\label{tab:exhaustive_10dB}
\begin{tabular}{cccccccccc}
\hline
\textbf{Rank \#} & $R$ (in bpcu) & $s_{1}$ & $s_{2}$ & $s_{3}$ & $s_{4}$ & $\rho_{1}$ & $\rho_{2}$ & $\rho_{3}$ & $\rho_{4}$  \\\hline\hline
\cellcolor{blue!25}1 & \cellcolor{blue!25}5.3992 & \cellcolor{blue!25}3   & \cellcolor{blue!25}3   &  \cellcolor{blue!25}4  & \cellcolor{blue!25}2   & \cellcolor{blue!25}0.27939 & \cellcolor{blue!25}0.28615 &  \cellcolor{blue!25}0.27362   & \cellcolor{blue!25}0.16083\\\hline
2   & 5.3952  &   3   &   3   &   3  &    3  &   0.27632   & 0.28255 &   0.27284   &  0.16828\\\hline
3  &  5.3927   &  3   &   3   &   5   &   1  &   0.29554  &  0.30523 &   0.30048   & 0.098749\\\hline
4   &   5.39  &   3   &   4    &  3  &    2  &   0.28069   & 0.27659  &   0.2789   & 0.16382\\\hline
 5 &   5.3882   &  3   &   5   &   2  &    2  &   0.28219   & 0.27762  &  0.27293   &  0.16725\\\hline
\end{tabular}
}
\end{table*}

\begin{figure*}[t]
  \centering
  \subfloat[]{
\includegraphics[width =.5\textwidth]{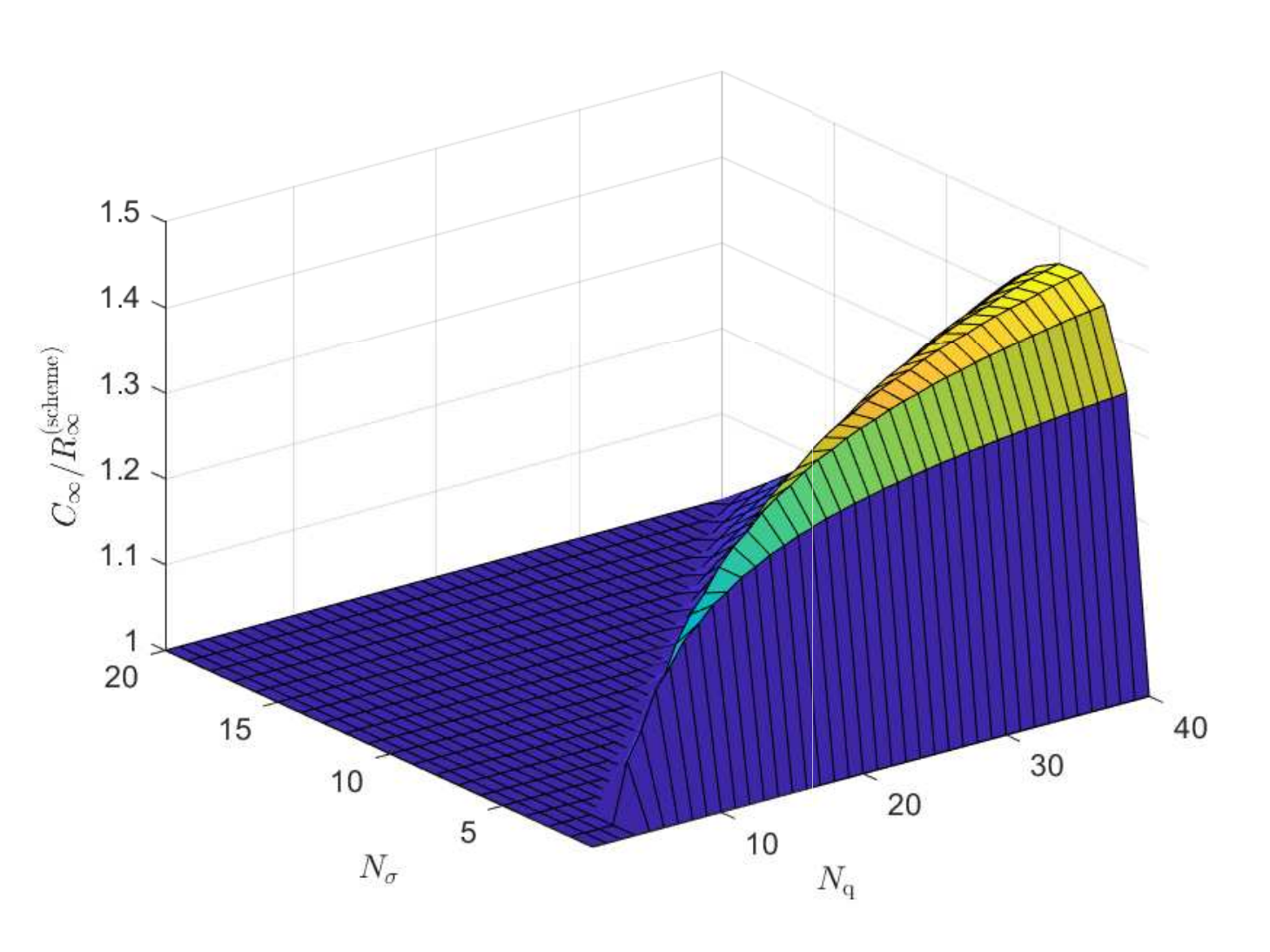}
\label{fig:converse_iso}
}
 \subfloat[]{
\includegraphics[width =0.5\textwidth]{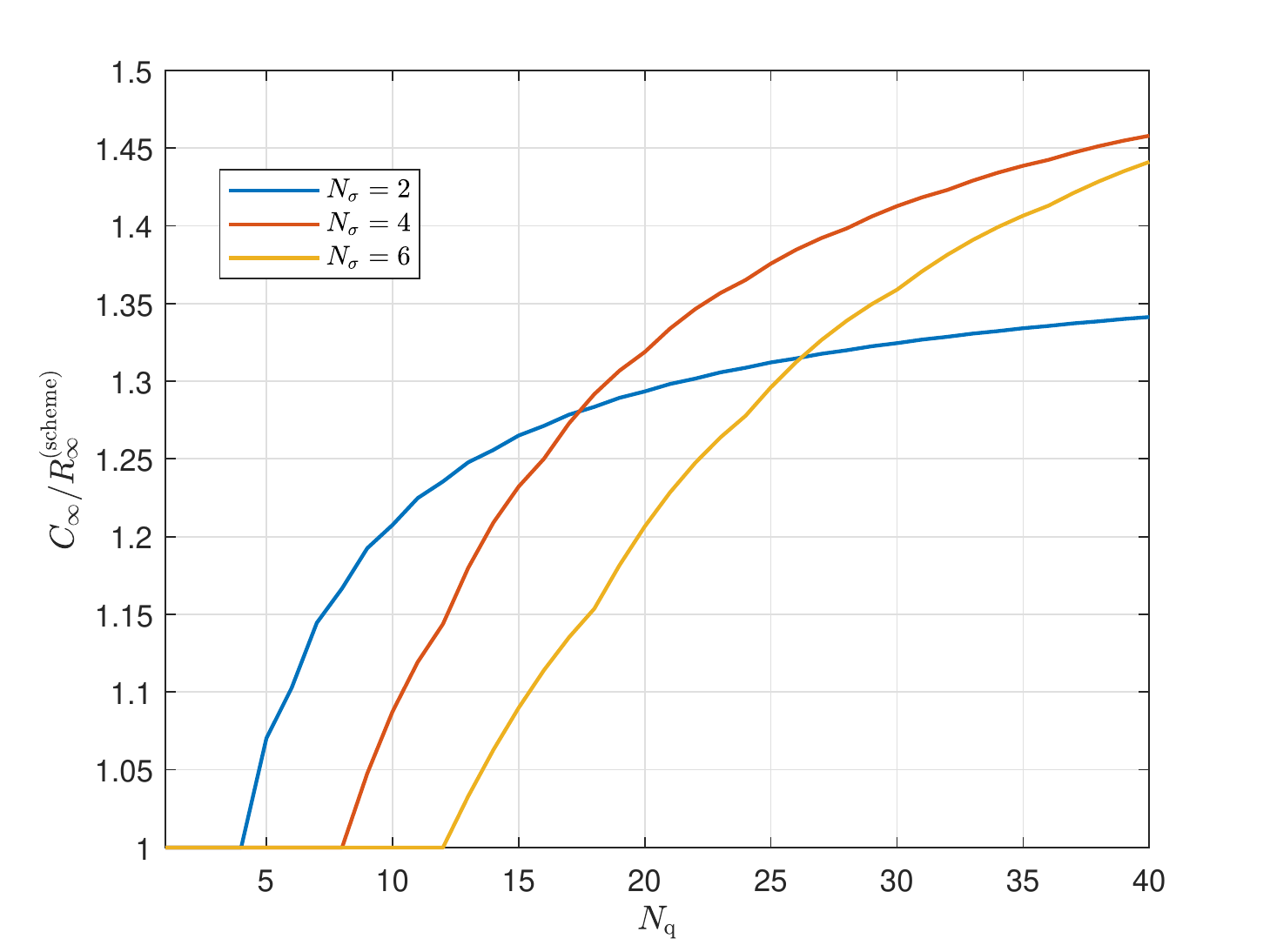}
    \label{fig:converse_1D}
} 

     \caption{(a) Ratio of $C_{\infty}/R_{\infty}^{(\mathrm{scheme})}$ as a function $N_{\sigma}$ and $N_{\mathrm{q}}$, and (b) plot of $C_{\infty}/R_{\infty}^{(\mathrm{scheme})}$ for $N_{\sigma} = 2,4,6$.}
     \label{fig:converse_ratio}
\end{figure*}

Despite the good agreement between our achievability scheme and the established capacity upper bound in the low SNR regime, the performance gap between the two widens as the SNR is increased in Figure \ref{fig:rate_vs_snr}. We shall refer to the rate of our scheme in the infinite SNR regime as $R_{\infty}^{(\mathrm{scheme})}$. $R_{\infty}^{(\mathrm{scheme})}$ is maximized if the quantizers are uniformly allocated to the $N_{\sigma}$ eigenchannels. In case $N_{\mathrm{q}}$ is not divisible by $N_{\sigma}$, we simply distribute the excess sign quantizers equally among the $N_{\mathrm{q}}\;\mathrm{mod}\; N_{\sigma}$ strongest eigenchannels. The number of possible outputs is
 \begin{align*}
     \mathcal{M}_{\mathrm{scheme}} =\begin{cases} \left\{2\left\lceil\frac{N_{\mathrm{q}}}{N_{\sigma}}\right\rceil\right\}^u\times \left\{2\left\lfloor\frac{N_{\mathrm{q}}}{N_{\sigma}}\right\rfloor\right\}^{N_\sigma - u}&,\; \left\lfloor\frac{N_{\mathrm{q}}}{N_{\sigma}}\right\rfloor > 0\\
     \left\{2\left\lceil\frac{N_{\mathrm{q}}}{N_{\sigma}}\right\rceil\right\}^u&,\; \mathrm{otherwise}
     \end{cases},
 \end{align*}
where $u = N_{\mathrm{q}}\;\mathrm{mod}\;N_{\sigma}$. The rate of our achievability scheme in the infinite SNR regime is $R^{(\mathrm{scheme})}_{\infty} = \log \mathcal{M}_{\mathrm{scheme}}$. To quantify the gap between $C_{\infty}$ and $R_{\infty}^{(\mathrm{scheme})}$ under different $N_{\mathrm{q}}$ and $N_{\sigma}$, we plot their ratio as a function of $N_{\sigma}$ and $N_{\mathrm{q}}$ in Figure \ref{fig:converse_iso}. It can be observed that $R_{\infty}^{(\mathrm{scheme})}$ coincides with $C_{\infty}$ when $2N_{\mathrm{q}} \leq N_{\sigma}$. This is the case in which we assign at most one sign quantizer to each real dimension. When $2N_{\mathrm{q}} > N_{\sigma}$, a logarithmic increase in $C_{\infty}/R_{\infty}^{(\mathrm{scheme})}$ is observed as $N_{\mathrm{q}}$ is increased. This is depicted in Figure \ref{fig:converse_1D}. The suboptimality of our achievability scheme in the infinite SNR regime comes from the restriction imposed on the matrix $\mathbf{\Phi}$. In our scheme, we simply design $\mathbf{\Phi}$ to connect a sign quantizer to a single eigenchannel and then choose the input distribution for each eigenchannel independently. On the other hand, $C_{\infty}$ is derived based on the intuition that a sign quantizer can be connected to multiple eigenchannels; thereby creating a hyperplane that passes through more than 2 real dimensions. In effect, the number of quantization regions created by $N_{\mathrm{q}}$ intersecting hyperplanes can exceed that of our achievability scheme.

\section{MIMO Receiver with Pipelined Phase ADC}
\label{section-MIMO_phaseADC}

To overcome the rate loss in the high SNR regime, we deviate our attention away from the one-shot receiver model and instead incorporate analog temporal and spatial processing techniques in the receiver design. To this end, we present a new MIMO receiver that utilizes an analog linear combiner and a more complex form of ADC structure. We call this ADC structure the \emph{pipelined phase ADC}, since the key idea is borrowed from the pipelined ADC topology \cite{Cho:1995}. In the subsequent discussion, we shall elaborate on the operation of this pipelined phase ADC. We then give a formal description of the new MIMO receiver in Section \ref{subsection-proposed_rx} and show how it can achieve the high SNR capacity of $N_{\mathrm{q}}$ bits/channel use.

\subsection{The Pipelined Phase ADC}
\begin{figure*}[t]
    \centering
    \subfloat[]{
    \includegraphics[scale =.55]{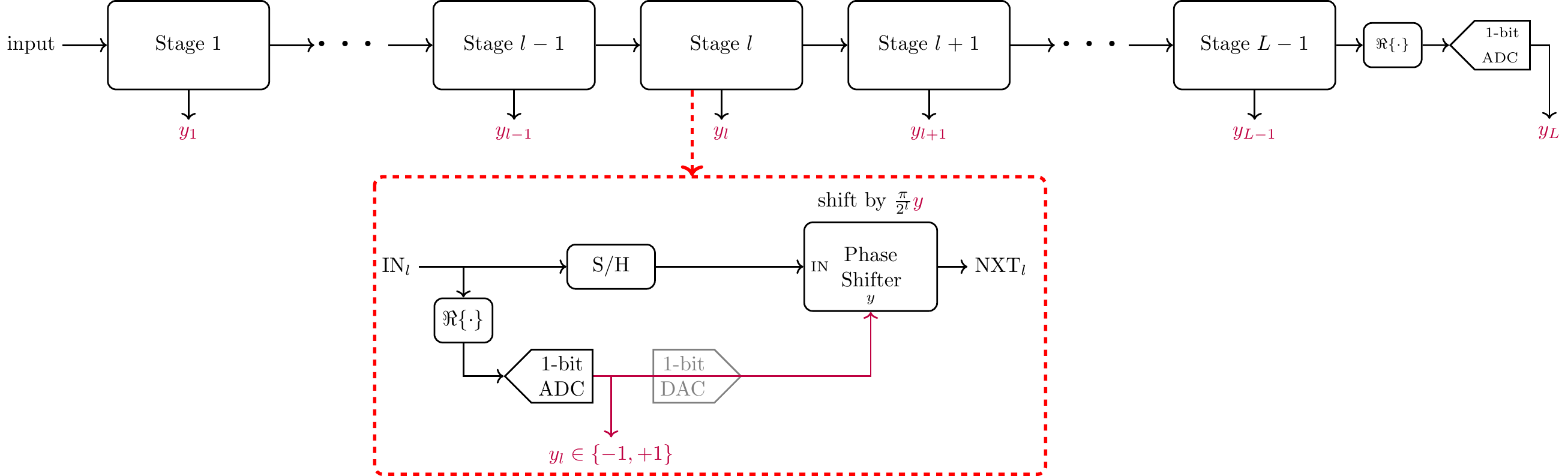}
    \label{fig:pipelined_PhaseADC}
    }\\
    \subfloat[]{
    \includegraphics[scale =.6]{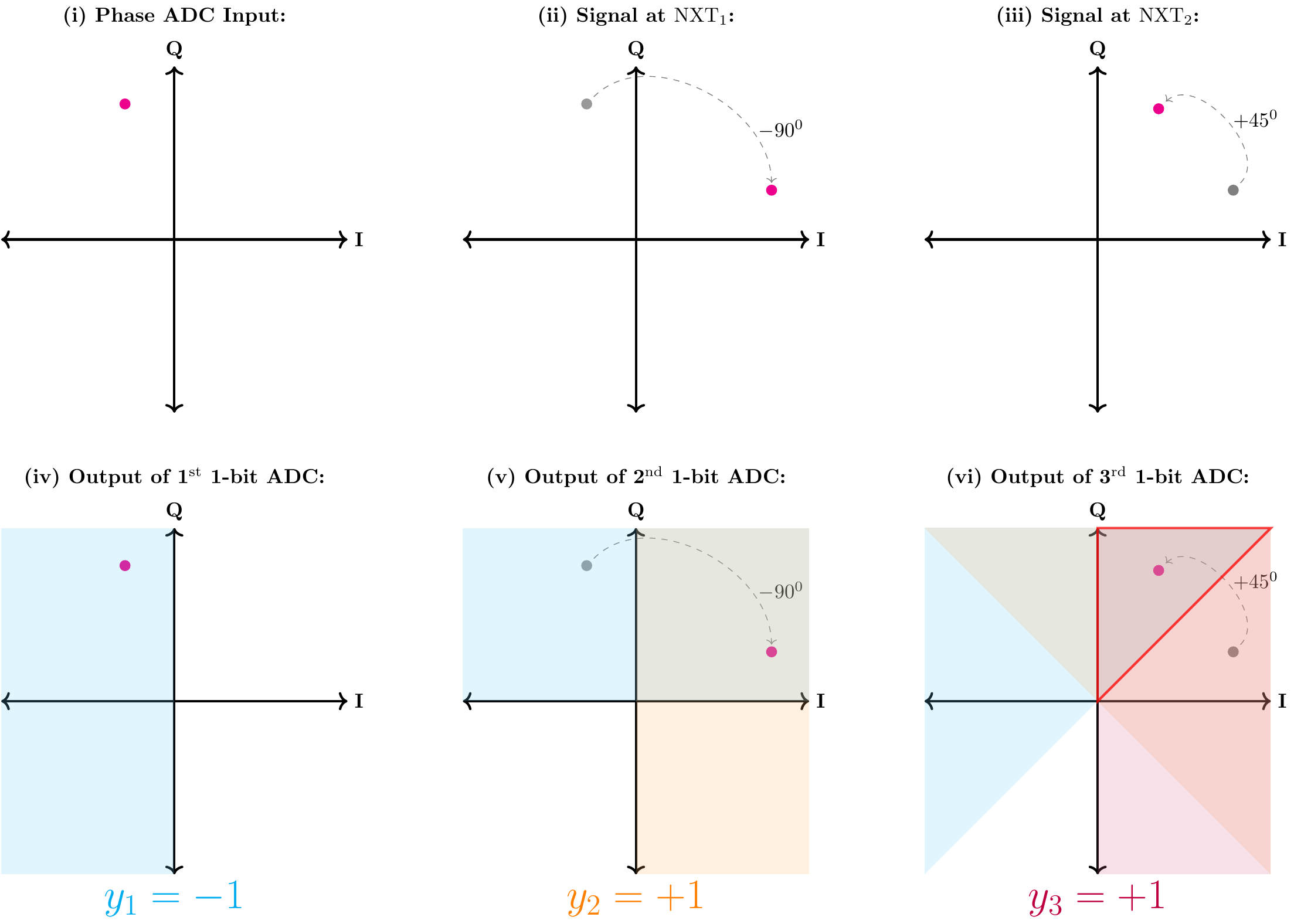}
    \label{fig:phaseADC_example}
    }
    \caption{(a) Block diagram of an $(L-1)$-stage Pipelined Phase ADC and (b) an illustrative example (for $L = 3$) to demonstrate the quantization process per stage.}
\end{figure*}

Figure \ref{fig:pipelined_PhaseADC} depicts a block diagram of the pipelined phase ADC. This ADC is composed of $L-1$ pipeline stages, where $L$ depends on the number of 1-bit ADCs. Each pipeline stage consists of an analog delay element, in the form of a sample-and-hold (S/H) block, and a phase shifter to perform analog temporal processing. At the $l$-th pipeline stage, we apply 1-bit quantization to the input to get $y_{l}$. The 1-bit output $y_{l}$ is then used by the phase shifter to apply an appropriate rotation to the S/H output\footnote{We note that a 1-bit digital-to-analog converter (DAC) might be required to interface the ADC output to the phase shifter.  However, the 1-bit DAC can be eliminated if the analog phase shifter is digitally-controlled.}. This phase shifted signal is then fed to the next pipeline stage for further processing. 

We give a more concrete example of the quantization mechanism through an illustrative example in Figure \ref{fig:phaseADC_example}. Here, we assume $L = 3$ so there are two pipeline stages and three 1-bit ADCs. The input signal is given by the magenta dot (shown in Figure \ref{fig:phaseADC_example}.i). The real component of the signal is fed to a 1-bit ADC of the 1$^{\mathrm{st}}$ pipelined stage to produce $y_{1}$ (Figure \ref{fig:phaseADC_example}.iv). Since it falls at the LHS of the $y$-axis (cyan region), the 1-bit ADC outputs $y_{1} = -1$. Consequently, this implies a clockwise phase shift of $\frac{\pi}{2} = 90^0$ (Figure \ref{fig:phaseADC_example}.ii); which will be fed to the next pipeline stage. This signal falls at the RHS of the $y$-axis (orange region); thus producing $y_{2} = +1$ (Figure \ref{fig:phaseADC_example}.v). Notice that the intersection of the cyan and orange regions forms a quantization region of a 2-bit phase quantizer. With $y_{2} = +1$, the signal at $\mathrm{NXT}_{2}$ is phase shifted by $\frac{\pi}{4} = 45^0$ counter clockwise. The last ADC outputs $y_{3} = +1$ since the resulting signal falls in the RHS of $y$-axis (purple region) (Figure \ref{fig:phaseADC_example}.vi). The intersection of the cyan, orange, and purple regions is a quantization region of a 3-bit phase quantizer. 

In general, the pipelined phase ADC enables us to create an $L$-bit phase quantizer with length-$L-1$ delay using $L$ 1-bit ADCs. Note that a flash ADC structure would require $2^L$ comparators to construct an $L$-bit phase quantizer. Moreover, because the analog pipelining structure enables each 1-bit ADC to extract 1-bit of information at each channel use, the maximum rate of $L$ bits/channel use is achievable. Note that the first $L - 1$ channel uses are strictly less than this rate since some analog delay elements do not contain signals initially. However, the definition of channel capacity applies for asymptotically large block lengths. Thus, this finite length delay is negligible in the asymptotic regime.

The proposed pipelined phase ADC has some resemblance with the ADC mechanism used in the adaptive threshold receiver recently proposed in \cite{Khalili:2021}. While both ADC topologies exploit analog domain pipelining, the latter adaptively chooses the locations of the 1-bit ADC thresholds in the current channel use based on the previous channel uses. The former applies an appropriate phase shift to the input of the next pipeline stage depending on the 1-bit ADC output in the current pipeline stage.

\subsection{Proposed Receiver}\label{subsection-proposed_rx}

\begin{figure*}[t]
    \centering
    \subfloat[]{
        \includegraphics[scale = .7]{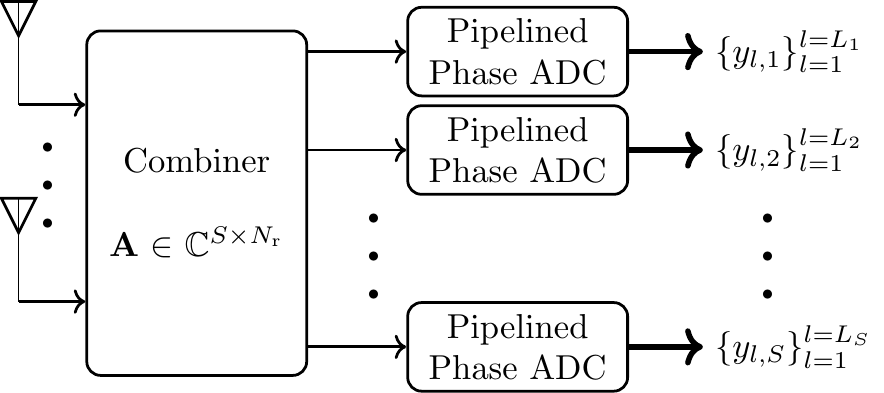}
        \label{fig:mimo_phaseADC}
    }
    \subfloat[]{
        \includegraphics[scale = .7]{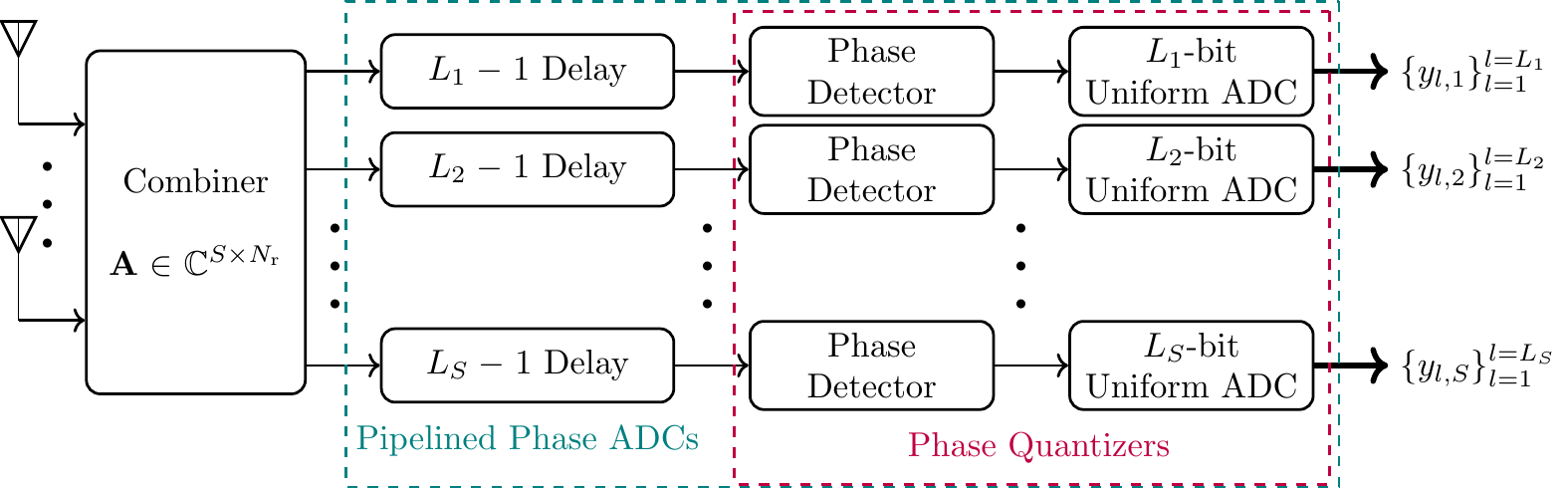}
        \label{fig:mimo_phaseADC_UB}
    }
    \caption{(a) System Model of MIMO Receiver with Pipelined Phase ADCs, and its (b) equivalent model}
\end{figure*}

The block diagram for the proposed MIMO receiver employing pipelined phase ADCs is shown in Figure \ref{fig:mimo_phaseADC}. An analog combiner is used to perform analog spatial processing of the received signals. The $S$ data streams at the analog combiner output are each fed to an $L_{S}$-bit pipelined phase ADC to produce $N_\mathrm{q} = \sum_{i = 1}^{S}L_{i}$ bits every channel use.

The achievability scheme presented in Section \ref{section-achievability_scheme} can be extended to this receiver structure. We consider the SVD of the channel, which gives a precoded transmit strategy $\tilde{\mathbf{x}} = \mathbf{V}\mathbf{x}$ and an analog linear combiner $\mathbf{A} = \mathbf{U}_{1}^H$. Thus, there are $S = N_{\mathrm{\sigma}}$ parallel data streams. Note that an $L$-bit pipelined phase ADC is equivalent to an $L$-bit phase quantizer (with some finite delay due to pipelining). As a result, the same reasoning in Section \ref{section-achievability_scheme} can be used to adapt \eqref{eq:achievability_opt} to this receiver architecture. The resulting optimization problem is
\begingroup
\allowdisplaybreaks
\begin{subequations}\label{eq:achievability_opt_modified}
\begin{alignat}{2}
\max_{L_{i},\rho_{i}} \quad &   \sum_{i=1}^{N_{\sigma}}L_{i} - w_{2^{L_{i}}}\left(\frac{\lambda_{i}\rho_{i}}{\sigma^2},\frac{\pi}{2^{L_{i}}}\right)\label{subeq:achievability_modified_objective}\\
\textrm{s.t.} \quad & \sum_{i = 1}^{N_{\sigma}}L_i = N_{\mathrm{q}}\\
  &\sum_{i = 1}^{N_{\sigma}}\rho_i \leq P\\
  &\qquad\rho_i \geq 0,\; L_{i}\in \{1,\cdots,N_{\mathrm{q}}\}
\end{alignat}
\end{subequations}
\endgroup
Consequently, Algorithm 1 can also be adapted to produce a heuristic solution to \eqref{eq:achievability_opt_modified} by modifying Lines 7 and 10 accordingly. 

As SNR grows unbounded, $w_{2^{L_i}}\left(\cdot,\frac{\pi}{2^{L_{i}}}\right)$ vanishes and  \eqref{subeq:achievability_modified_objective} approaches $N_{\mathrm{q}}$ bits/channel use. This is the maximum rate that any channel with $N_{\mathrm{q}}$-bit output quantization can achieve. Note that this rate can be larger than $C_{\infty}$ established in Section \ref{subsection-cap_upperbound_Inf}. This is because the inclusion of analog temporal processing allows an analog sample to be quantized multiple times; thus the combinatorial geometry approach used in Section \ref{subsection-cap_upperbound_Inf} should be modified accordingly. To this end, we simply use the trivial upper bound $N_{\mathrm{q}}$ bit/channel use.

For the capacity upper bound in the finite SNR regime, we can use the DPI argument in Section \ref{subsection-cap_upperbound_finite} to show that $C_{\phi-\mathrm{detector}}$ upper bounds the capacity of our proposed MIMO receiver. The following corollary of Proposition \ref{proposition:finite_SNR_upperbound} extends this result.

\begin{corollary}
The capacity of the Gaussian channel employing the MIMO receiver with pipelined phase ADCs can be upper bounded by \eqref{eq:capacity_UB}.
\end{corollary}
\begin{proof}
To prove the claim, we consider the equivalent receiver model in Figure \ref{fig:mimo_phaseADC_UB}. By DPI, the capacity is bounded by the maximum mutual information between the transmitted symbols and the output of the phase detector. Moreover, the finite length delay prior to the phase detector does not change the capacity.
\end{proof}

We also point out that the use of sample-and-hold blocks, 1-bit digital-to-analog converters (DACs), and analog phase shifters in the pipelined phase ADCs entails additional power consumption to the MIMO receiver. In this work, we simply focus at how analog spatial and temporal processing can be used to maximize the achievable rate of a MIMO receiver for a given number of 1-bit ADCs. Determining the best architecture from an energy efficiency standpoint can be a future direction of this study.

\section{Numerical Results for the MIMO Receiver with Pipelined Phase ADCs} \label{section-numerical_result2}

\begin{figure*}[t]
  \centering
 \subfloat[]{
\includegraphics[width =0.525\textwidth]{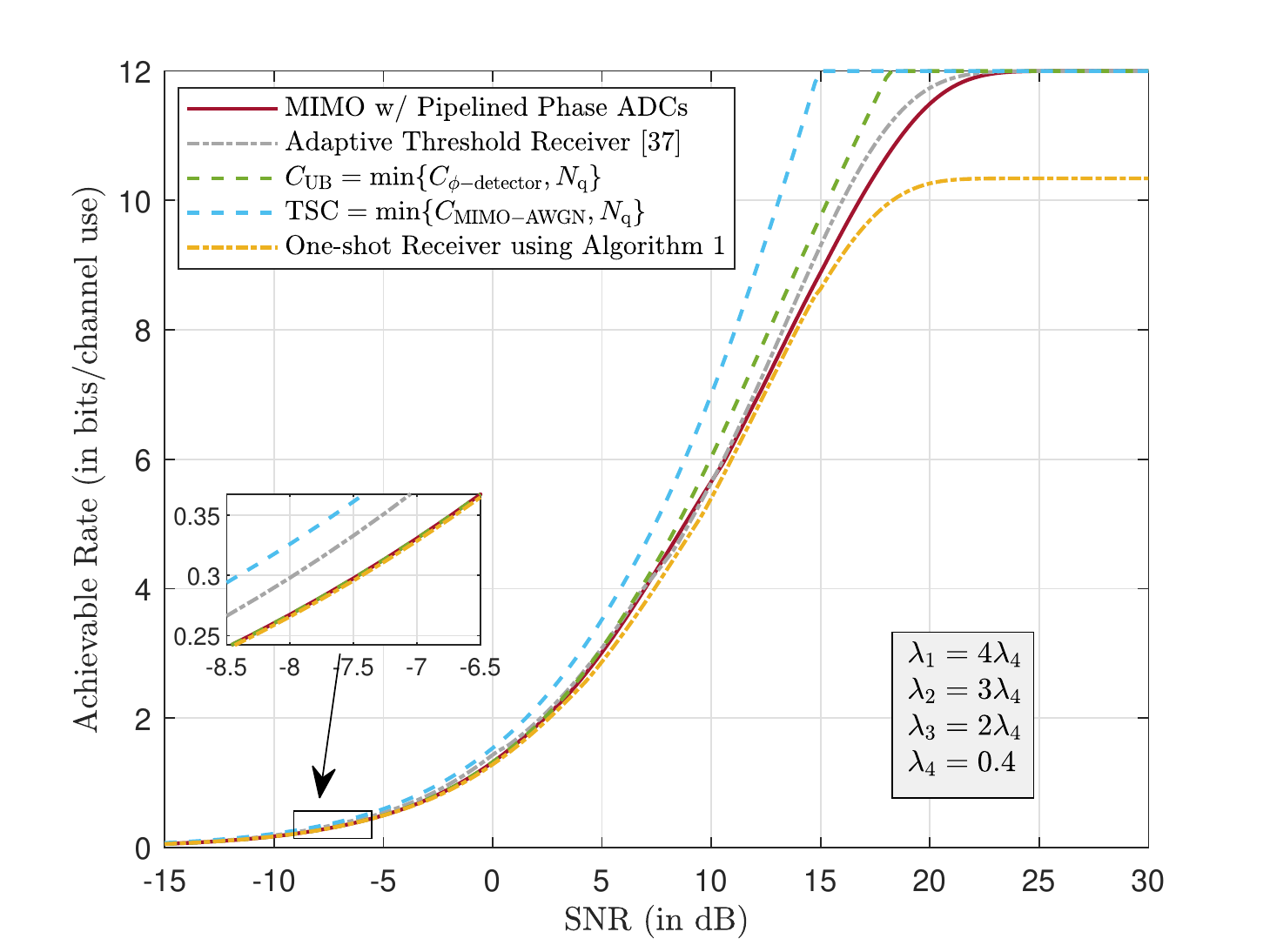}
    \label{fig:rate_vs_snr_pipelined_lambda1234}
}
    \hspace*{-.75cm}%
     \subfloat[]{
\includegraphics[width =.525\textwidth]{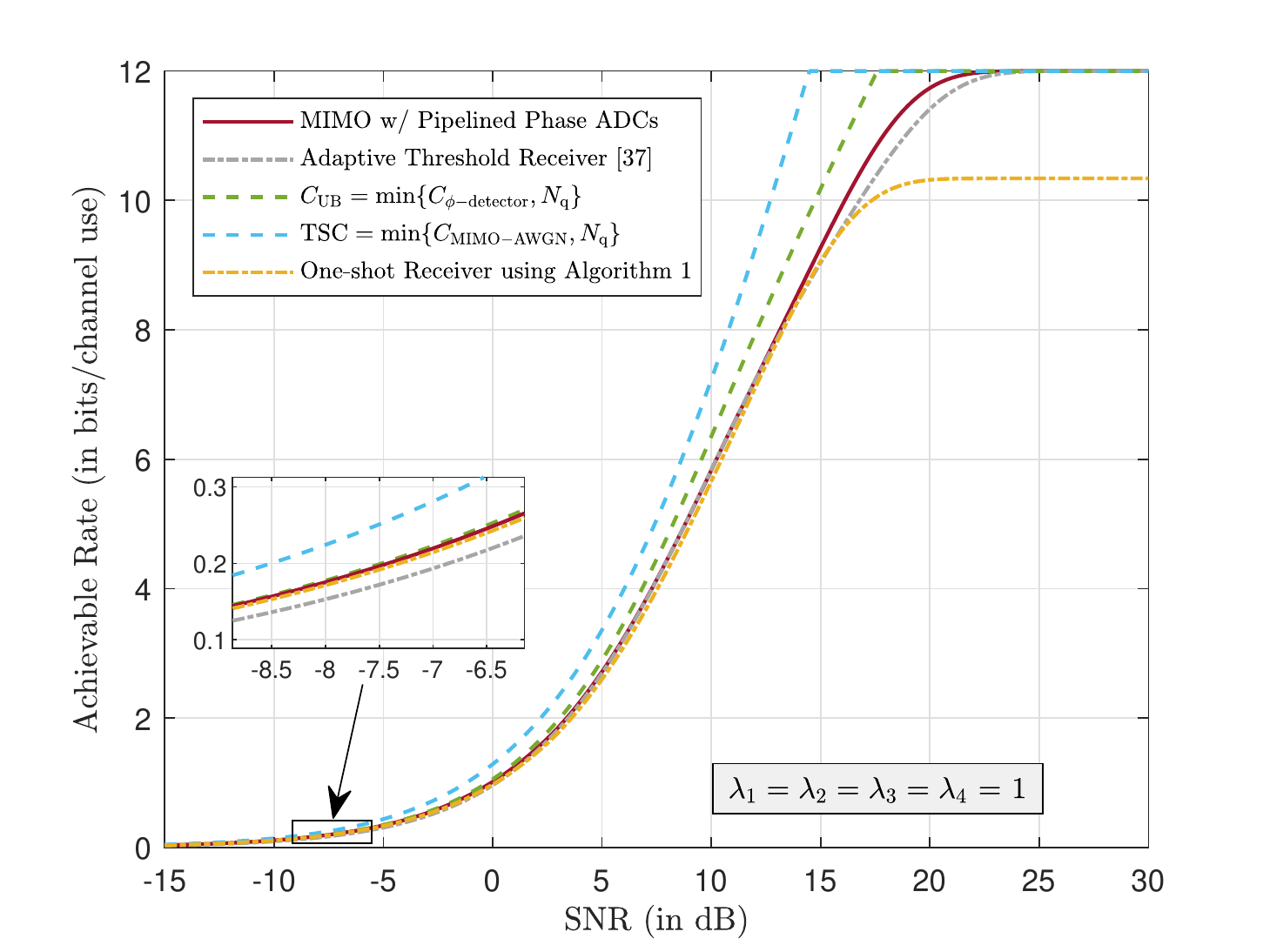}
\label{fig:rate_vs_snr_pipelined_EqualLambda}
}
     \caption{ Achievable Rate vs. SNR of the Proposed Receiver for (a) $\lambda_1 = 1.6, \lambda_2 = 1.2, \lambda_3 = 0.8,$ and $\lambda_4 = 0.4$; and  (b) $\lambda_1 = \lambda_2 = \lambda_3 = \lambda_4 = 1$ ($N_{\mathrm{q}} = 12$). The rate is compared to that of the adaptive threshold receiver.}
     \label{fig:proposed_receiver_rate}
\end{figure*}

In this section, we examine the achievable rate of the MIMO reciever with pipelined phase ADCs. A 4$\times$4 MIMO Gaussian channel with $N_{\mathrm{q}} = 12$ available 1-bit ADCs is considered. We set $P = 1$ and vary the SNR by changing the noise variance $\sigma^2$. Moreover, we look at two sets of eigenvalues for the experiment setup: (a) $\lambda_1 = 1.6, \lambda_2 = 1.2, \lambda_3 = 0.8,$ and $\lambda_4 = 0.4$; and (b) $\lambda_1 = \lambda_2 = \lambda_3 = \lambda_4 = 1$. We shall refer to these channel setups as setup A and setup B.

The achievable rates of the MIMO receiver employing pipelined phase ADCs for the two for setup A and setup B are depicted in Figure \ref{fig:rate_vs_snr_pipelined_lambda1234} and \ref{fig:rate_vs_snr_pipelined_EqualLambda}, respectively. To compute the achievable rate, we modify Algorithm 1 as described in Section \ref{section-MIMO_phaseADC}. We also superimpose the achievable rate of the hybrid one-shot receiver with zero threshold ADCs, the capacity upper bound described in Section \ref{section-MIMO_phaseADC}, and the TSC bound. It can be observed that the achievability scheme for the MIMO receiver with pipelined phase ADCs is tight with our established upper bound in the low SNR regime and also attains the high SNR capacity of $N_{\mathrm{q}}$ bits/channel. While the gap between the achievable rate of MIMO receiver with pipelined phase ADCs and that of the hybrid one-shot receiver in Section \ref{section-achievability_scheme} is small in the low SNR regime, this gap gradually increases with SNR. This demonstrates that the rate increase provided by incorporating analog temporal processing in our receiver design is more pronounced in the high SNR regime.

We compare the performance of our proposed receiver to that of the adaptive threshold receiver in \cite{Khalili:2021}. Since the adaptive threshold receiver is designed for real MIMO channels, we made some modifications in the channel setup for fair comparison. We considered an 8$\times$8 real MIMO Gaussian channel with eigenvalues $\lambda'_{2i-1} = \lambda'_{2i} = \lambda_{i}$ for $i = 1,2,3,4$. Furthermore, we set $\mathrm{SNR} = 2P/\sigma^2$ instead of $\mathrm{SNR} = P/\sigma^2$. The adaptive threshold receiver effectively creates parallel real eigenchannels with uniform quantization at the output. The transmit strategy for the acheivability scheme described in \cite{Khalili:2021} is to send equiprobable pulse amplitude modulation (PAM) over each eigenchannel. The power allocation per PAM strategy is obtained using the conventional waterfilling algorithm for the unquantized AWGN channel. Using this power allocation scheme, an exhaustive search procedure is performed to allocate the 1-bit ADCs. The achievable rate results in Figure \ref{fig:rate_vs_snr_pipelined_lambda1234} and \ref{fig:rate_vs_snr_pipelined_EqualLambda} show that the achievable rate of our proposed receiver may have inferior or superior performance than the adaptive threshold receiver depending on the SNR and channel eigenvalues. 

One potential reason why our proposed receiver works better in some cases is because the transmit power and the 1-bit ADC allocation are jointly optimized. This is in contrast to the adaptive threshold receiver which performs separate optimization of the transmit power and ADC allocation. On the other hand, the adaptive threshold receiver extracts the amplitude information, which is neglected by our proposed receiver. This may explain why the adaptive threshold receiver outperformed our proposed receiver in Figure \ref{fig:rate_vs_snr_pipelined_lambda1234}. Nonetheless, we point out that the adaptive threshold receiver requires AGCs to adjust the dynamic range of the received signal.

\section{Conclusion}
\label{section-conclusion}

In this work, we analyzed the capacity of a point-to-point Gaussian MIMO channel in which the receiver is equipped with $N_{\mathrm{q}}$ 1-bit ADCs and an analog linear combiner prior to quantization. In particular, we focused on the zero-threshold ADC case. Our first contribution is an achievability scheme in which the analog combiner is configured to create parallel Gaussian channels with phase quantization at the output. The achievable rate of this constructed channel is evaluated using an alternating optimization approach. We then etablished a new capacity upper bound that is tighter than the TSC bound when the ADCs are restricted to have zero threshold. This upper bound serves as a measure of the worst-case gap between the rate of our achievability scheme and the capacity of the channel. Our numerical results showed that the rate of our achievability scheme is tight in the low SNR regime. However, a performance gap exists in the high SNR regime whenever $N_{\sigma} \leq 2N_{\mathrm{q}}$. More precisely, when this condition is satisfied, we observed that the ratio of the channel capacity and the rate of our achievability scheme in the infinite SNR regime grows logarithmically with the number of 1-bit ADCs. To overcome this, a new receiver is proposed that implements joint analog spatial and temporal processing through the use of an analog combiner and pipelined phase ADCs. We showed that the proposed receiver achieves the high SNR capacity of $N_{\mathrm{q}}$ bits/channel use and outperforms the adaptive threshold receiver \cite{Khalili:2021} when the channel eigenvalues are equal. Further research needs to be conducted to be able to generalize these results to multi-user setting and different fading environments. As mentioned in Section \ref{section-MIMO_phaseADC}, investigation of the best architecture from an energy efficiency viewpoint is another interesting research direction.

\begin{appendices}
\section{Proof of Lemma \ref{lemma:active_Ns}}\label{proof_Ns}

Suppose the optimal strategy $\mathcal{O}'$ uses the ordered set of eigenchannels with eigenvalues \[\mathcal{S}' = \{\lambda_1,\cdots,\lambda_{N_{\mathrm{s}}+1}\}\backslash\lambda_i\]
for some integer $i\in\{1,\cdots,N_{\mathrm{s}}\}$. The power and sign quantizers allocated to the $k$-th eigenchannel in this optimal strategy $\mathcal{O}'$ are denoted as $\rho_k'$ and $s_{k}'$, respectively. Define another strategy $\mathcal{O}^*$ which uses the ordered set of eigenchannels with eigenvalues $\mathcal{S}^* = \{\lambda_1,\cdots,\lambda_{N_{\mathrm{s}}}\}$. Let $\rho_k^*$ and $s_k^*$ be the power and sign quantizer allocation in the $k$-th eigenchannel when the strategy $\mathcal{O}^*$ is used. If we set $\rho_k^* = \rho_k'$ and $s_k^* = s_k'$  $\forall k = 1,\cdots,i-1,i+1,\cdots, N_{\mathrm{s}}$ and let $\rho_i^* = \rho_{N_{\mathrm{s}}+1}'$ and $s_i^* = s_{N_{\mathrm{s}}+1}'$, then the difference between the rate of $\mathcal{O}^*$ and $\mathcal{O}'$ is
\begin{align*}
    R_{\mathcal{O}^*} -  R_{\mathcal{O}} &= w_{2s_{i}}\left(\frac{\lambda_{N_{\mathrm{s}}+1}\rho_{i}^*}{\sigma^2},\frac{\pi}{2s_{i}}\right) -w_{2s_{i}}\left(\frac{\lambda_{i}\rho_{i}^*}{\sigma^2},\frac{\pi}{2s_{i}}\right) \\
    &\geq 0.
\end{align*}
The inequality comes from the monotonic decreasing property of phase quantization entropy with respect to $\nu$ \cite[Proposition 1]{bernardo2021TIT} and $\lambda_{N_\mathrm{s}+1} \leq \lambda_{i}$. This contradicts the assumption that strategy $\mathcal{O}'$ is optimal.

\section{Proof of Correctness of the Dynamic Programming Approach}\label{proof_dp}

The key technique in showing the correctness is through strong induction. First, the base case $i = 0$ or $n_{\mathrm{q}} = 0$ is true since if there is no channel to send information or there is no sign quantizer that can output produce the output $\mathbf{y}$, then $I(\mathbf{x};\mathbf{y})$ should be 0. Next, we prove the inductive step. Assume $f(i',n'_{\mathrm{q}})$ be the optimal solution  for all $i' < i$. We need to show  $f(i,n_{\mathrm{q}})$ is the optimal solution for the state $(i,n_{\mathrm{q}})$. Note that the sum capacity of channels $1$ to $i$ with $n_{\mathrm{q}}$ available sign quantizers can be expressed as
\begin{align*}
    &\sum_{j=1}^{i}\log 2 s_{j} - w_{2s_{j}}\left(\frac{\lambda_{j}\rho_{j}}{\sigma^2},\frac{\pi}{2s_{j}}\right) \\
    &\qquad= \sum_{j=1}^{i-1}\log 2 s_{j} - w_{2s_{j}}\left(\frac{\lambda_{j}\rho_{j}}{\sigma^2},\frac{\pi}{2s_{j}}\right) \\
    &\qquad\quad+ \log 2 s_{i} - w_{2s_{i}}\left(\frac{\lambda_{i}\rho_{i}}{\sigma^2},\frac{\pi}{2s_{i}}\right)\\
    &\qquad\leq\; f(i-1,n_{\mathrm{q}} - s_{i}) + \log 2 s_{i} - w_{2s_{i}}\left(\frac{\lambda_{i}\rho_{i}}{\sigma^2},\frac{\pi}{2s_{i}}\right).
\end{align*}
The first line follows from isolating the capacity of the $i$-th channel from channels $i$ to $i - 1$. The inequality in the second line follows from the optimality of $f(i',n_{\mathrm{q}}')$ and equality is achieved by choosing the optimal $s_{i}$. Hence, the problem has an optimal substructure property. The algorithm considers all possible choices of $s_{i}$  and compares their values. Thus, optimality of $f(i,n_{\mathrm{q}})$ is guaranteed.

\section{Proof of Lemma \ref{lemma:reduce_branches}}\label{proof_reduce_branches}

By the chain rule of mutual information, we have
\begin{align*}
    I_{\mathbf{A}}(\mathbf{x};\tilde{\mathbf{y}}) = & I_{\mathbf{A}}(\mathbf{x};\tilde{\mathbf{y}}^{(1)},\tilde{\mathbf{y}}^{(2)})\\
    =& I_{\mathbf{A}_1}(\mathbf{x};\tilde{\mathbf{y}}^{(1)}) + I_{\mathbf{A}_2}(\mathbf{x};\tilde{\mathbf{y}}^{(2)}|\tilde{\mathbf{y}}^{(1)}).
\end{align*}
The claim is proven if we can show that $I_{\mathbf{A}_2}(\mathbf{x};\tilde{\mathbf{y}}^{(2)}|\tilde{\mathbf{y}}^{(1)}) = 0$. In other words, $\mathbf{x} \rightarrow \tilde{\mathbf{y}}^{(1)} \rightarrow \tilde{\mathbf{y}}^{(2)}$ should form a Markov chain. The term $I_{\mathbf{A}_2}(\mathbf{x};\tilde{\mathbf{y}}^{(2)}|\tilde{\mathbf{y}}^{(1)})$ can be expressed as 
\begingroup
\allowdisplaybreaks
\begin{align*}
    =& I_{\mathbf{A}_2}\left(\mathbf{x};   e^{j\tilde{\mathbf{y}}^{(2)}}\;\big|\;  {\tilde{\mathbf{y}}^{(1)}}\right)\\
    =& I_{\mathbf{A}_2}\left(\mathbf{x};   e^{j\left\{\mathbf{z}_{N_{\sigma}+1:N_{\mathrm{q}}}' - \mathbf{B}^{(2)}\{\mathbf{B}^{(1)}\}^{\dagger}\mathbf{z}_{1:N_{\sigma}}'\right\}}\;\bigg|\;  {\tilde{\mathbf{y}}^{(1)}}\right)\\
    =&  I_{\mathbf{A}_2}\left(\mathbf{x};\mathbf{z}_{N_{\sigma}+1:N_{\mathrm{q}}}' - \mathbf{B}^{(2)}\{\mathbf{B}^{(1)}\}^{\dagger}\mathbf{z}_{1:N_{\sigma}}'\;\big|\;\tilde{\mathbf{y}}^{(1)}\right)\\
    =& 0.
\end{align*}
\endgroup
The equality in the first line follows from the fact that $e^{j(\cdot)}$ is bijective. The second equality is obtained by noting that
 \begin{align*}
     &\exp\left(j  \tilde{\mathbf{y}}^{(2)}\right)\cdot\exp\left(-j \mathbf{B}^{(2)}\{\mathbf{B}^{(1)}\}^{\dagger}  \tilde{\mathbf{y}}^{(1)}\right)\\ &\qquad\qquad= \exp\left(j(\tilde{\mathbf{y}}^{(2)} - \mathbf{B}^{(2)}\{\mathbf{B}^{(1)}\}^{\dagger}  \tilde{\mathbf{y}}^{(1)})\right)\\
     &\qquad\qquad= \exp\left(j(\mathbf{z}_{N_{\sigma}+1:N_{\mathrm{q}}}' - \mathbf{B}^{(2)}\{\mathbf{B}^{(1)}\}^{\dagger}\mathbf{z}_{1:N_{\sigma}}')\right),
 \end{align*}
 where $\{\cdot\}^{\dagger}$ is the Moore-Penrose inverse operator. Since the transmitted symbol and the phase of the additive noise components are independent, $I_{\mathbf{A}_2}(\mathbf{x};\tilde{\mathbf{y}}^{(2)}|\tilde{\mathbf{y}}^{(1)}) = 0$.
 

To prove the second claim, we note that we have full control over $\mathbf{A}$. If $\mathbf{B}_1$ is not full rank, we can create $\mathbf{A}'$ by permuting the rows of $\mathbf{A}$ to make $\mathbf{B}_1'$ full rank. This, in effect, reorders the elements of $\tilde{\mathbf{y}}$ but does not change the mutual information since reordering is a bijective mapping. Thus, $I_{\mathbf{A}}(\mathbf{x};\tilde{\mathbf{y}}) = I_{\mathbf{A}'}(\mathbf{x};\tilde{\mathbf{y}})$. If no such row permutation of rows can make $\mathbf{B}_1$ full rank, then this implies that $\mathrm{rank}\{\mathbf{A}\mathbf{H}\} < N_{\sigma}$. We can simply use an analog combiner $\mathbf{A}'$ with $\mathrm{rank\{\mathbf{A}'\mathbf{H}\}} = N_{\sigma}$ and employ a transmit strategy $F_{\mathbf{X}'}$ that only uses the $\mathrm{rank}\{\mathbf{A}\mathbf{H}\}$ out of the $N_{\sigma}$ eigenchannels so that $I_{\mathbf{A}}(\mathbf{x};\tilde{\mathbf{y}}) = I_{\mathbf{A}'}(\mathbf{x}';\tilde{\mathbf{y}})$.

\end{appendices}

\ifCLASSOPTIONcaptionsoff
  \newpage
\fi

\bibliographystyle{ieeetr}
\bibliography{references}

\begin{IEEEbiography}[{\includegraphics[width=1in,height=1.25in,clip,keepaspectratio]{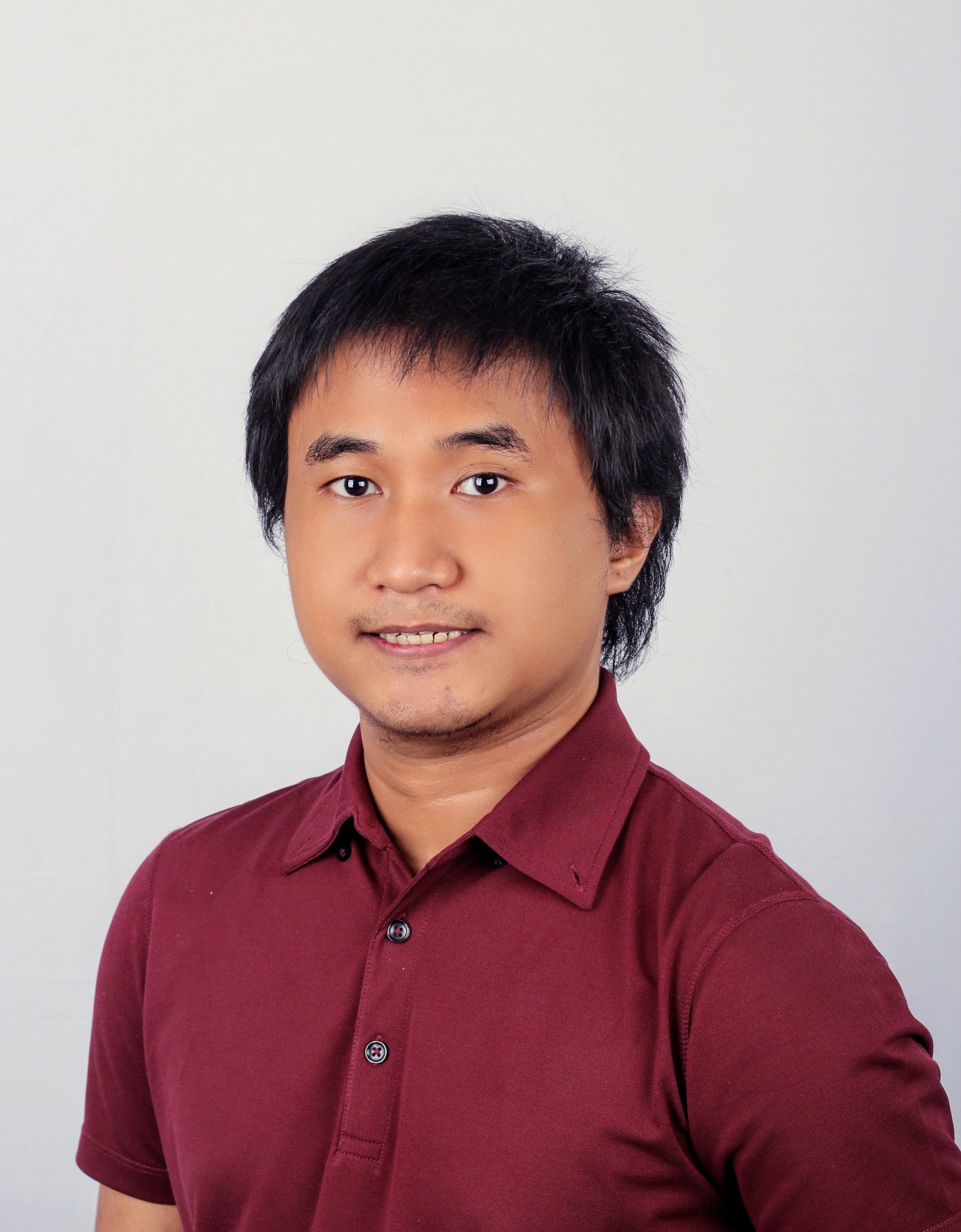}}]{Neil Irwin Bernardo} (Graduate Student Member, IEEE) received the B.S. degree in electronics and communications engineering and the M.S. degree in electrical engineering from the University of the Philippines Diliman in 2014 and 2016, respectively. He is currently pursuing the Ph.D. degree in engineering with The University of Melbourne, Australia. He has been a faculty member of the University of the Philippines Diliman since 2014. His research interests include wireless communications, signal processing, and information theory.
\end{IEEEbiography}
\vskip 0pt plus -1fil
\begin{IEEEbiography}[{\includegraphics[width=1in,height=1.25in,clip,keepaspectratio]{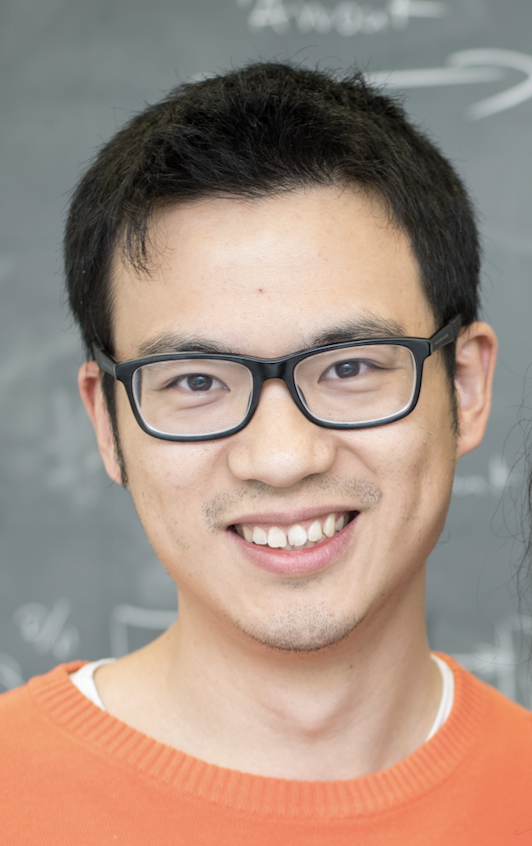}}]{Jingge Zhu} (Member, IEEE) received the B.S. degree and M.S. degree in electrical engineering from Shanghai Jiao Tong University, Shanghai, China, in 2008 and 2011, respectively, the Dipl.-Ing. degree in technische Informatik from Technische Universit\"{a}t Berlin, Berlin, Germany in 2011 and the Doctorat \`{e}s Sciences degree from the Ecole Polytechnique F\'{e}d\'{e}rale (EPFL), Lausanne, Switzerland, in 2016. He was a post-doctoral researcher at the University of California, Berkeley from 2016 to 2018.  He is now a lecturer at the University of Melbourne, Australia. His research interests include information theory with applications in communication systems and machine learning. 

Dr. Zhu received the Discovery Early Career Research Award (DECRA) from the Australian Research Council in 2021, the IEEE Heinrich Hertz Award for Best Communications Letters in 2013, the Early Postdoc. Mobility Fellowship from Swiss National Science Foundation in 2015, and the Chinese Government Award for Outstanding Students Abroad in 2016.
\end{IEEEbiography}
\vskip 0pt plus -1fil
\begin{IEEEbiography}[{\includegraphics[width=1in,height=1.25in,clip,keepaspectratio]{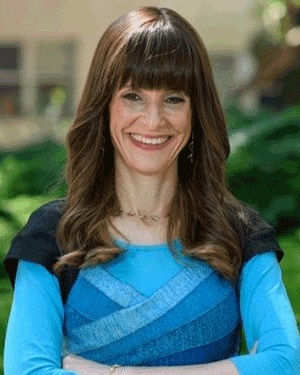}}]{Yonina C. Eldar} (Fellow, IEEE) received the B.Sc. degree in Physics in 1995 and the B.Sc. degree in Electrical Engineering in 1996 both from Tel-Aviv University (TAU), Tel-Aviv, Israel, and the Ph.D. degree in Electrical Engineering and Computer Science in 2002 from the Massachusetts Institute of Technology (MIT), Cambridge. She is currently a Professor in the Department of Mathematics and Computer Science, Weizmann Institute of Science, Rehovot, Israel. She was previously a Professor in the Department of Electrical Engineering at the Technion. She is also a Visiting Professor at MIT, a Visiting Scientist at the Broad Institute, and an Adjunct Professor at Duke University and was a Visiting Professor at Stanford. She is a member of the Israel Academy of Sciences and Humanities (elected 2017), an IEEE Fellow and a EURASIP Fellow. Her research interests are in the broad areas of statistical signal processing, sampling theory and compressed sensing, learning and optimization methods, and their applications to biology, medical imaging and optics.

Dr. Eldar has received many awards for excellence in research and teaching, including the  IEEE Signal Processing Society Technical Achievement Award (2013), the IEEE/AESS Fred Nathanson Memorial Radar Award (2014), and the IEEE Kiyo Tomiyasu Award (2016). She was a Horev Fellow of the Leaders in Science and Technology program at the Technion and an Alon Fellow. She received the Michael Bruno Memorial Award from the Rothschild Foundation, the Weizmann Prize for Exact Sciences, the Wolf Foundation Krill Prize for Excellence in Scientific Research, the Henry Taub Prize for Excellence in Research (twice), the Hershel Rich Innovation Award (three times), the Award for Women with Distinguished Contributions, the Andre and Bella Meyer Lectureship, the Career Development Chair at the Technion, the Muriel \& David Jacknow Award for Excellence in Teaching, and the Technion’s Award for Excellence in Teaching (two times).  She received several best paper awards and best demo awards together with her research students and colleagues including the SIAM outstanding Paper Prize, the UFFC Outstanding Paper Award, the Signal Processing Society Best Paper Award and the IET Circuits, Devices and Systems Premium Award, was selected as one of the 50 most influential women in Israel and in Asia, and is a highly cited researcher.

She was a member of the Young Israel Academy of Science and Humanities and the Israel Committee for Higher Education. She is the Editor in Chief of Foundations and Trends in Signal Processing, a member of the IEEE Sensor Array and Multichannel Technical Committee and serves on several other IEEE committees. In the past, she was a Signal Processing Society Distinguished Lecturer, member of the IEEE Signal Processing Theory and Methods and Bio Imaging Signal Processing technical committees, and served as an associate editor for the IEEE Transactions On Signal Processing, the EURASIP Journal of Signal Processing, the SIAM Journal on Matrix Analysis and Applications, and the SIAM Journal on Imaging Sciences. She was Co-Chair and Technical Co-Chair of several international conferences and workshops. She is author of the book "Sampling Theory: Beyond Bandlimited Systems" and co-author of five other books published by Cambridge University Press.
\end{IEEEbiography}
\vskip 0pt plus -1fil
\begin{IEEEbiography}[{\includegraphics[width=1in,height=1.25in,clip,keepaspectratio]{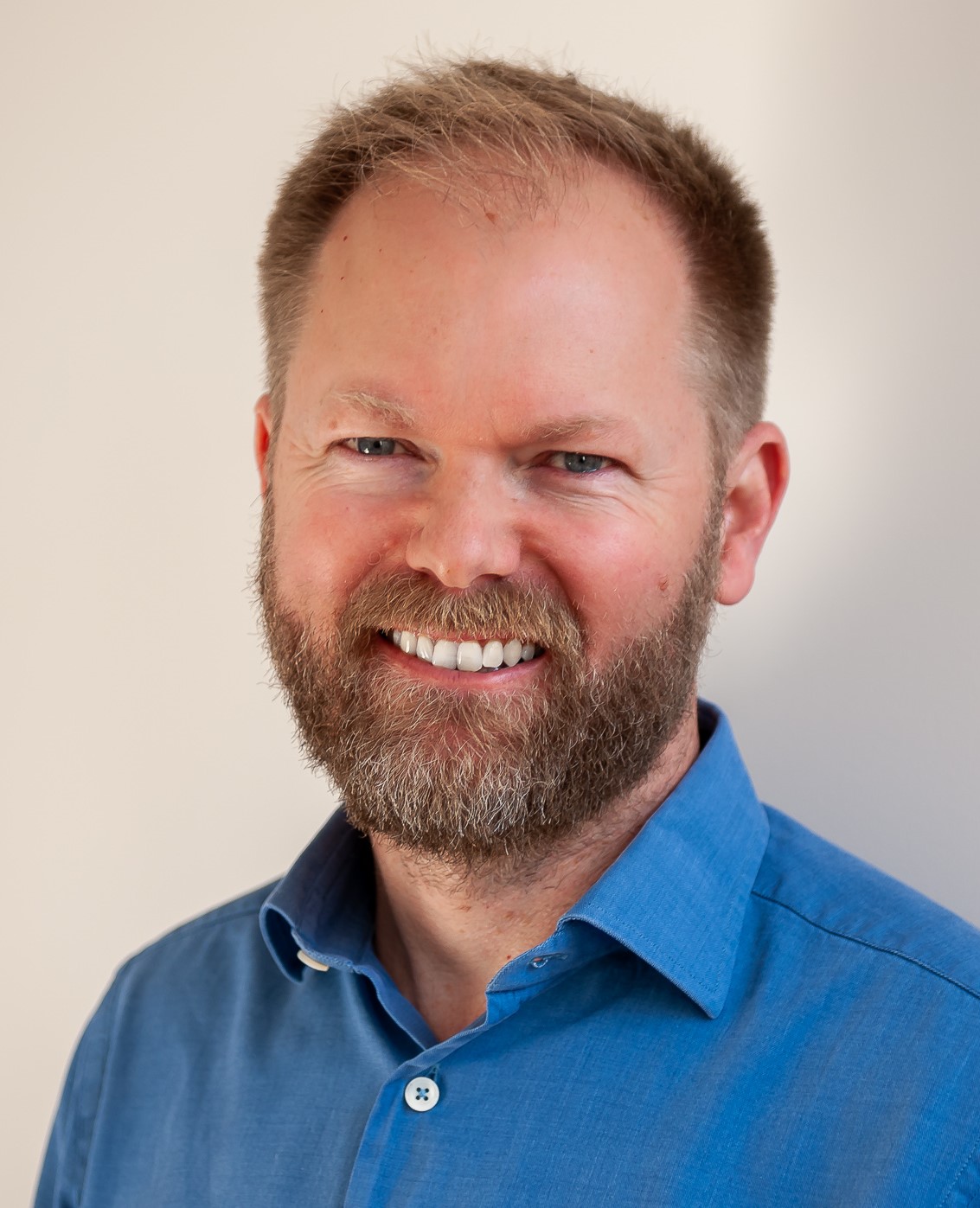}}]{Jamie Evans} (Senior Member, IEEE) was born in Newcastle, Australia, in 1970. He received the B.S. degree in physics and the B.E. degree in computer engineering from the University of Newcastle, in 1992 and 1993, respectively, where he received the University Medal upon graduation. He received the M.S. and the Ph.D. degrees from the University of Melbourne, Australia, in 1996 and 1998, respectively, both in electrical engineering, and was awarded the Chancellor's Prize for excellence for his Ph.D. thesis. From March 1998 to June 1999, he was a Visiting Researcher in the Department of Electrical Engineering and Computer Science, University of California, Berkeley. Since returning to Australia in July 1999 he has held academic positions at the University of Sydney, the University of Melbourne and Monash University. He is currently a Professor of Electrical and Electronic Engineering and Pro Vice-Chancellor (Education) at the University of Melbourne. His research interests are in communications theory, information theory, and statistical signal processing with a focus on wireless communications networks.
\end{IEEEbiography}

\end{document}